\RequirePackage{fix-cm}
\documentclass[twocolumn]{svjour3}          % twocolumn
\smartqed  % flush right qed marks, e.g. at end of proof
\usepackage{graphicx}
\usepackage{subfigure}
\usepackage{mathtools}
\usepackage{url}
\usepackage{multirow}
\usepackage{mdframed}

\begin{document}

\title{An analytical framework to nowcast well-being using mobile phone data%\thanks{Grants or other notes
}


\author{Luca Pappalardo *$\dagger$         \and Maarten Vanhoof $\ddagger$ \and Lorenzo Gabrielli $\dagger$ \and Zbigniew Smoreda $\ddagger$ \and Dino Pedreschi * \and Fosca Giannotti $\dagger$
}


\institute{*Department of Computer Science \at
University of Pisa, Italy \\
\email{lpappalardo@di.unipi.it}
\and
$\dagger$Institute of Information Science and Technologies (ISTI) \at
National Research Council (CNR), Italy \\
\email{fosca.giannotti@isti.cnr.it}           %  \\
\and
$\ddagger$SENSE \at
Orange Labs, France\\
\email{zbigniew.smoreda@orange.com}  
}

\date{}

\maketitle

\begin{abstract}
An intriguing open question is whether measurements made on Big Data recording human activities can yield us high-fidelity proxies of socio-economic development and well-being. Can we monitor and predict the socio-economic development of a territory just by observing the behavior of its inhabitants through the lens of Big Data? In this paper, we design a data-driven analytical framework that uses mobility measures and social measures extracted from mobile phone data to estimate indicators for socio-economic development and well-being. We discover that the {\em diversity of mobility}, defined in terms of entropy of the individual users' trajectories, exhibits (i) significant correlation with two different socio-economic indicators and (ii) the highest importance in predictive models built to predict the socio-economic indicators. Our analytical framework opens an interesting perspective to study human behavior through the lens of Big Data by means of new statistical indicators that quantify and possibly ``nowcast'' the well-being and the socio-economic development of a territory.
\end{abstract}

\section{Introduction}
Big Data, the masses of digital breadcrumbs produced by the information technologies that humans use in their daily activities, allow us to scrutinize individual and collective behavior at an unprecedented scale, detail, and speed. Building on this opportunity we have the potential capability of creating a digital nervous system of our society, enabling the measurement, monitoring and prediction of relevant aspects of the socio-economic structure in quasi real time \cite{Giannotti}. An intriguing question is whether and how measurements made on Big Data can yield us high-fidelity proxies of socio-economic development and well-being. Can we monitor and possibly predict the socio-economic development of our societies just by observing human behavior, for example human movements and social relationships, through the lens of Big Data?

This fascinating question, also stimulated by the United Nations in recent reports \cite{worldcount,onureport}, has attracted the interest of researchers from several disciplines, who started investigating the relations between human behavior and economic development based on large experimental datasets collected for completely different purposes \cite{eagle,gabrielli2015}. As a first result along this line a seminal work exploited a nationwide mobile phone dataset to discover that the diversity of social contacts of the inhabitants of a municipality is positively associated to a socio-economic indicator of poverty, independently surveyed by the official statistics institutes \cite{eagle}. This result suggests that social behavior, to some extent, is a proxy for the economic status of a given territory. However, little effort has been put in investigating how \emph{human mobility} affects, and is affected by, the socio-economic development of a territory. Theoretical works suggest that human mobility is related to economic well-being, as it could nourish economic and facilitate flows of people and goods, whereas constraints in the possibilities to move freely can diminish economic opportunities \cite{kwan99}. So, it is reasonable to investigate the role of human mobility with respect to the socio-economic development of a given territory. 

Our paper provides a twofold contribution. First, we design a data-driven analytical framework that uses Big Data to extract meaningful measures of human behavior and estimate indicators for the socio-economic development. The analytical framework we propose is repeatable on different countries and geographic scales since it is based on mobile phone data, the so-called CDR (Call Detail Records) of calling and texting activity of users. Mobile phone data, indeed, can be retrieved in every country due to their worldwide diffusion \cite{blondel2015survey}: there are 6.8 billion mobile phone subscribers today over 7 billion people on the planet, with  a penetration of 128\% in the developed world and 90\% in developing countries. CDR data have proven to be a hi-fi proxy for individuals' movements and social interactions \cite{gonzalez08,onnela07}. 

Second, we apply the analytical framework on large-scale mobile phone data and quantify the relations between human mobility, social interactions and economic development in France using municipality-level official statistics as external comparison measurements. We first define four individual measures over mobile phone data which describe different aspects of individual human behavior: the volume of mobility, the diversity of mobility, the volume of sociality and the diversity of sociality. Each individual measure is computed for each of the several million users in our dataset based on their locations and calls as recorded in the mobile phone data.
In a second stage, we aggregate the four individual measures at the level of French municipalities and explore the correlations between the four aggregated measures and two external indicators of socio-economic development. We find that the aggregated mobility diversity of individuals resident in the same municipality exhibits a superior correlation degree with the socio-economic indicators and we confirm these results against two different null models, an observation that allows us to reject the hypothesis that our discovery occurred by chance.

Next, we build regression and classification models to predict the external socio-economic indicators from the population density and the social and mobility measures aggregated at municipality scale. We show that the {\em diversity of human mobility} adds a significant predictive power in both regression and classification models, far larger than the diversity of social contacts and demographic measures such as population density, a factor that is known to be correlated with the intensity of human activities \cite{pan2012,yan2013}. The importance of this finding is twofold. On one side, it offers a new stimulus to social research: diversity is a key concept not only for natural ecosystems but also for the social ecosystems, and can be used to understand deeply the complexity of our interconnected society. On the other side, our results reveal the high potential of Big Data in providing representative, relatively inexpensive and readily available measures as proxies of economic development. Our analytical framework opens an interesting perspective to engineer official statistics processes to monitor human behavior through mobile phone data. New statistical indicators can be defined to describe and possibly ``nowcast'' the economic status of a territory, even when such measurements would be impossible using traditional censuses and surveys \cite{worldcount,onureport}.

\begin{figure*}[htb!]
\begin{center}
	\resizebox*{17.6cm}{!}{\includegraphics{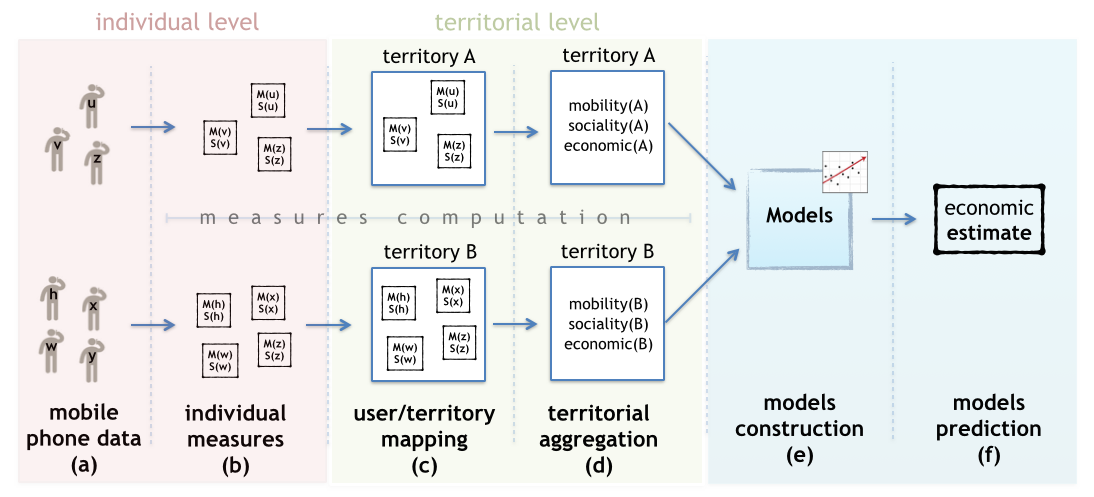}}
\caption{The data-driven analytical framework. Starting from mobile phone data (a) mobility and social measures are computed for each individual in the dataset (b). Each individual is then assigned to the territory where she resides (c) and the individual measures are aggregated at territorial level (d). Starting from the aggregated measures predictive models are constructed (e) in order to estimate and predict the socio-economic development of the territories (f).}
\label{fig:process}
\end{center}
\end{figure*}

The paper is organized as follows. Section \ref{sec:related} revises the scientific literature relevant to our topic, Section \ref{sec:process} describes in detail the data-driven analytical framework we propose. In Section \ref{sec:experimental}, Section \ref{sec:analysis}  and Section \ref{sec:models} we apply our analytical framework on a nationwide mobile phone dataset covering several weeks of call activity in France. We introduce the mobile phone data in Section \ref{sec:data}, the measures of individual mobility behavior and individual social behavior in Section \ref{sec:measures}, and the computations of the measures on a nation-wide mobile phone dataset in Section \ref{sec:computation}. In Section \ref{sec:analysis} we describe the results of the correlation analysis and validate them against two null models. In Section \ref{sec:models} we present and validate predictive models for socio-economic development. In Section \ref{sec:discussion} we discuss the results and finally Section \ref{sec:conclusions} concludes the paper describing the opportunities and the challenges that arise from our research.

\section{Related work}
\label{sec:related}
The interest around the analysis of Big Data and the possibility to compile them into a comprehensive picture of human behavior have infected all branches of human knowledge, from sports \cite{dsaa15} to economy \cite{pennacchio}. However, two aspects in particular attracted the interest of scientists in the last decade, due the striking abundance of data in those contexts: human mobility and social networks. 

Studies from different disciplines document a stunning heterogeneity of human travel patterns \cite{gonzalez08,pappalardo2013}, and at the same time observe a high degree of predictability \cite{song2010limits,pentland}. The patterns of human mobility have been used to build generative models of individual human mobility and human migration flows \cite{karamshuk2011human,simini}, methods for profiling individuals according to their mobility patterns \cite{pappalardo2015}, to discover geographic borders according to recurrent trips of private vehicles \cite{borders}, or to predict the formation of social ties \cite{cho,wang2011}, and classification models to predict the kind of activity associated to individuals' trips on the only basis of the observed displacements \cite{Liao2007,gonzalez_clustering,rinzi14}. In the context of social network analysis the observation of social interactions data provided by emails, mobile phones, and social media allowed to reveal the complexity underlying the social structure \cite{barabba}: hubs exist in our social networks who strongly contribute to the so-called small world phenomenon \cite{Backstrom:2012:FDS:2380718.2380723}, and social networks are found to have a tendency to partition into social communities, i.e.\ clusters of densely connected sets of individuals \cite{Fortunato201075}. 

The last few years have also witnessed a growing interest around the usage of Big Data to support official statistics in the measurement of individual and collective well-being \cite{daas,daas2}. Even the United Nations, in two recent reports, stimulate the usage of Big Data to investigate the patterns of phenomena relative to people's health and well-being \cite{worldcount}\cite{onureport}. The vast majority of works in the context of Big Data for official statistics are based on the analysis of mobile phone data, the so-called CDR (Call Detail Records) of calling and texting activity of users. Mobile phone data, indeed, guarantee the repeatability of experiments on different countries and geographical scales since they can be retrieved nowadays in every country due to their worldwide diffusion \cite{blondel2015survey}. A set of recent works use mobile phone data as a proxy for socio-demographic variables.  
Deville et al., for example, show how the ubiquity of mobile phone data can be exploited to provide accurate and detailed maps of population distribution over national scales and any time period \cite{deville}. Brea et al.\ study the structure of the social graph of mobile phone users of Mexico and propose an algorithm for the prediction of the age of mobile phone users \cite{brea}. Another recent work uses mobile phone data to study inter-city mobility and develop a methodology to detect the fraction of residents, commuters and visitors within each city \cite{sis2014}.

A lot of effort has been put in recent years on the usage of mobile phone data to study the relationships between human behavior and collective socio-economic development. The seminal work by Eagle et al.\ analyzes a nationwide mobile phone dataset and shows that, in the UK, regional communication diversity is positively associated to a socio-economic ranking \cite{eagle}. Gutierrez et al.\ address the issue of mapping poverty with mobile phone data through the analysis of airtime credit purchases in Ivory Coast \cite{journals/corr/GutierrezKB13}. Blumenstock shows a preliminary evidence of a relationship between individual wealth and the history of mobile phone transactions \cite{blumenstock}. Decuyper et al.\ use mobile phone data to study food security indicators finding a strong correlation between the consumption of vegetables rich in vitamins and airtime purchase \cite{DBLP:journals/corr/DecuyperRWBKGBL14}. Frias-Martinez et al.\ analyze the relationship between human mobility and the socio-economic status of urban zones, presenting which mobility indicators correlate best with socio-economic levels and building a model to predict the socio-economic level from mobile phone traces \cite{Frias-martinez_1can}. Pappalardo et al.\ analyze mobile phone data and extract meaningful mobility measures for cities, discovering interesting correlation between human mobility aspects and socio-economic indicators \cite{pappalardo_bigdata}.
Lotero et al.\ analyze the architecture of urban mobility networks in two Latin-American cities from the multiplex perspective. They discover that the socio-economic characteristics of the population have an extraordinary impact in the layer organization of these multiplex systems \cite{lotero2014several}. Amini et al.\ use mobile phone data to compare human mobility patterns of a developing country (Ivory Coast) and a developed country (Portugal). They show that cultural diversity in developing regions can present challenges to mobility models defined in less culturally diverse regions \cite{amini2014}. Smith-Clarke at al.\ analyze the aggregated mobile phone data of two developing countries and extract features that are strongly correlated with poverty indexes derived from official statistics census data \cite{poverty_cheap}.
Other recent works use different types of mobility data, e.g.\ GPS tracks and market retail data, to show that Big Data on human movements can be used to support official statistics and understand people's purchase needs. Pennacchioli et al.\ for example provide an empirical evidence of the influence of purchase needs on human mobility, analyzing the purchases of an Italian supermarket chain to show a range effect of products: the more sophisticated the needs they satisfy, the more the customers are willing to travel \cite{pennacchioli2013}. Marchetti et al.\ perform a study on a regional level analyzing GPS tracks from cars in Tuscany to extract measures of human mobility at province and municipality level, finding a strong correlation between the mobility measures and a poverty index independently surveyed by the Italian official statistics institute \cite{gabrielli2015}. 

Despite an increasing interest around this field, a view on the state-of-the-art cannot avoid to notice that there is no a unified methodology to exploit Big Data for official statistics. It is also surprising that widely accepted measures of human mobility (e.g.\ radius of gyration \cite{gonzalez08} and mobility entropy \cite{song2010limits}) have not been used so far. We overcome these issues by providing an analytical framework as support for official statistics, which allows for a systematic evaluation of the relations between relevant aspects of human behavior and the development of a territory. Moreover, our paper shows how standard mobility measures, not exploited so far, are powerful tools for official statistics purposes.

\section{The Analytical Framework}
\label{sec:process}

Our analytical framework is a knowledge and analytical infrastructure that uses Big Data to provide reliable measurements of socio-economic development, aiming at satisfying the increasing demand by policy makers for continuous and up-to-date information on the geographic distribution of poverty, inequality or life conditions. Figure \ref{fig:process} describes the structure of the methodology we propose. The analytical framework is based on mobile phone data, which guarantee the repeatability of the process on different countries and geographical scales. Mobile phone data are indeed ubiquitous and can be retrieved in every country due to their worldwide diffusion: nowadays the penetration of mobile phones is of 128\% in developed countries and 90\% in developing countries, with 6.8 billion mobile phone subscribers today over 7 billion people on the planet \cite{blondel2015survey}. In particular the call detail records (CDR), generally collected by mobile phone operators for billing and operational purposes, contain an enormous amount of information on how, when, and with whom people communicate. This wealth of information allows to capture different aspects of human behavior and stimulated the creativity of scientists from different disciplines, who demonstrated that mobile phone data are a high quality proxy for studying individual mobility and social ties \cite{gonzalez08,onnela07}.

Starting from the collected mobile phone data (Figure \ref{fig:process}(a)) a set of measures are computed which grasp the salient aspects of individuals' mobility and social behavior (Figure \ref{fig:process}(b)). This step is computationally expensive when the analytical framework is applied on massive data such as the CDRs of an entire country for a long period. To parallelize the computations and speed up the execution a distributed processing platform can be used such as Hadoop or Spark. A wide set of mobility and social measures can be computed during this phase, and the set can be enlarged with new measures as soon as they are proven to be correlated with socio-economic development aspects of interest. In Section \ref{sec:measures} we propose, as an example, a set of standard measures of individual mobility and sociality and show how they can be computed on mobile phone data.

As generally required by policy makers, official statistics about socio-economic development are available at the level of geographic units, e.g.\ regions, provinces, municipalities, districts or census cells. Therefore, the individuals in the dataset have to be mapped to the corresponding territory of residence, in order to perform an aggregation of the individual measures into a territorial measure (Figure \ref{fig:process}(c) and \ref{fig:process}(d)). When the city of residence or the address of the users are available in the data, this information can be easily used to assign each individual to corresponding city of residence. Unfortunately these socio-demographic data are generally not available in mobile phone data for privacy and proprietary reasons. This issue can be solved, with a certain degree of approximation, by inferring the information from the data source. In literature the phone tower where a user makes the highest number of calls during nighttime is usually considered her home phone tower \cite{smoreda2012}. Then with standard Geographic Information System techniques it is possible to associate the phone tower to its territory (see Section \ref{sec:computation}). 

The obtained aggregated measures are compared with the external socio-economic indicators to perform correlation analysis, learn and evaluate predictive models (Figure \ref{fig:process}(e)). The predictive models can be aimed at predicting the actual value of socio-economic development of the territory, e.g.\ by regression models (Section \ref{sec:regressions}), or to predict the class of socio-economic development, i.e.\ the level of development of a given geographic unit as done by classification models (Section \ref{sec:classifications}). Finally, the estimates and the predictions produced by the models are the output of the analytical framework (Figure \ref{fig:process}(f)). The measures, the territorial aggregation and the predictive models of the analytical framework can be updated every time new mobile phone data become available, providing policy makers with up-to-date estimates of the socio-economic situation of a given territory, in contrast with indicators produced by official statistics institutes with are generally released after months or even once a year.

In the following sections we apply the proposed analytical framework on a large-scale nation-wide mobile phone dataset and describe its implementation step by step: from the definition of measures on the data (Sections \ref{sec:data} and \ref{sec:measures}), to their computation and territorial aggregation (Section \ref{sec:computation}), and the construction of predictive models (Sections \ref{sec:correlations}, \ref{sec:regressions} and \ref{sec:classifications}).

\section{Measuring Human Behavior}
\label{sec:experimental}
We now discuss steps (a), (b) and (c) in Figure \ref{fig:process}, presenting the experimental setting which consists in the computation of the individual measures on the data and their aggregation at territorial level. First, we describe the mobile phone data we use as proxy for individual behavior, together with details about data preprocessing (Section \ref{sec:data}). Then we define the individual measures capturing diverse aspects of individual mobility and social behavior (Section \ref{sec:measures}). Finally we show how we compute the individual measures and aggregate them at municipality level (Section \ref{sec:computation}). 

\subsection{Mobile phone data}
\label{sec:data}
We have access to a set of Call Detail Records (CDR) gathered for billing purposes by Orange mobile phone operator, recording 215 million calls made during 45 days by 20 million anonymized mobile phone users.
CDRs collect geographical, temporal and interaction information on mobile phone use and show an enormous potential to empirically investigate human dynamics on a society wide scale \cite{Hidalgo2008}. Each time an individual makes a call the mobile phone operator registers the connection between the caller and the callee, the duration of the call and the coordinates of the phone tower communicating with the served phone, allowing to reconstruct the user's time-resolved trajectory. Table \ref{tab:CDR_data} illustrates an example of the structure of CDRs.

\begin{table}[htb]\centering
\def\arraystretch{1.1}
\subfigure[]{
\normalsize
\begin{tabular}{| c | c | c | c |}
\hline
\textbf{timestamp} & \textbf{tower} & \textbf{caller} & \textbf{callee} \\
\hline
2007/09/10 23:34 & 36 & 4F80460 & 4F80331\\
2007/10/10 01:12 & 36 & 2B01359 & 9H80125\\
2007/10/10 01:43 & 38 & 2B19935 & 6W1199\\
\vdots & \vdots & \vdots & \vdots \\
\hline
\end{tabular}
}
\subfigure[]{
\normalsize
\begin{tabular}{| r | r | r |}
\hline
\textbf{tower} & \textbf{latitude} & \textbf{longitude}\\
\hline
36 & 49.54 & 3.64\\
37 & 48.28 & 1.258\\
38 & 48.22 & -1.52\\
\vdots & \vdots & \vdots \\
\hline
\end{tabular}
}
\caption{Example of Call Detail Records (CDRs). Every time a user makes a call, a record is created with timestamp, the phone tower serving the call, the caller identifier and the callee identifier (a). For each tower, the latitude and longitude coordinates are available to map the tower on the territory (b).}
\label{tab:CDR_data}
\end{table}

\begin{figure}[htb!]
\begin{center}
	\resizebox*{7cm}{!}{\includegraphics{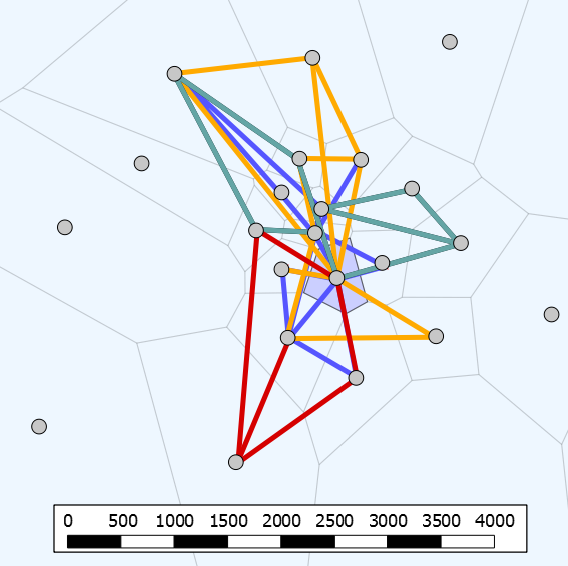}}
\caption{The detailed trajectory of a single user. The phone towers are shown as grey dots, and the Voronoi lattice in grey marks approximate reception area of each tower. CDRs records the identity of the closest tower to a mobile user; thus, we can not identify the position of a user within a Voronoi cell. The trajectory describes the user's movements during 4 days (each day in a different color). The tower where the user made the highest number of calls during nighttime is depicted in bolder grey.}
\label{fig:example_trajs}
\end{center}
\end{figure}

In order to focus on individuals with reliable statistics, we carry out some preprocessing steps. First, we select only users with a call frequency higher than the threshold $f=N/45>0.5$, where $N$ is the number of calls made by the user and 45 days is the length of our period of observation, we delete all the users with less than one call every two days (in average over the observation period). 

\begin{figure}[htb!]
\begin{center}
	\resizebox*{7cm}{!}{\includegraphics{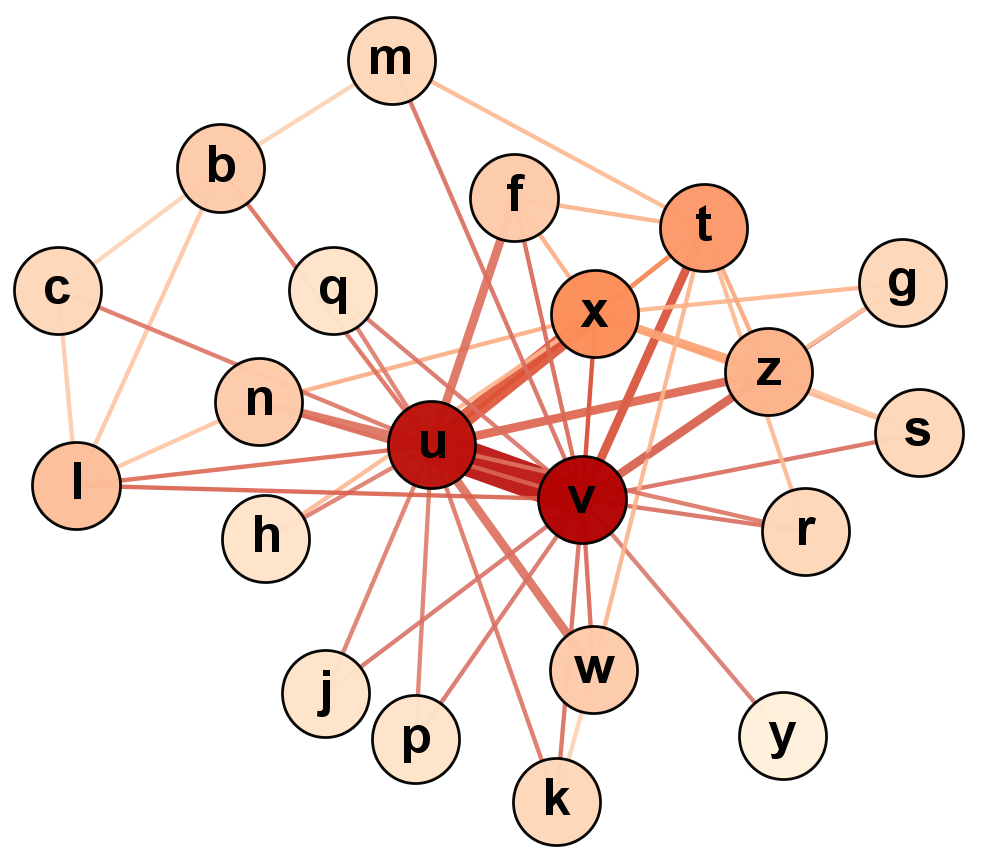}}
\caption{A fraction of the call graph centered on a single user $u$. Nodes represent users, edges indicate reciprocated calls between the users, the size of the edges is proportional to the total number of calls between the users during the 45 days.}
\label{fig:example_network}
\end{center}
\end{figure}

Second, we reconstruct the mobility trajectories and the social network of the filtered users. We reconstruct the trajectory of a user based on the time-ordered list of cell phone towers from which she made her calls during the period of observation (see Figure \ref{fig:example_trajs}).
We then translate the CDR data into a social network representation by linking two users if at least one reciprocated pair of calls exists between them during the period of observation (i.e.\ A called B and B called A). This procedure eliminates a large number of one-way calls, most of which correspond to isolated events and do not represent meaningful communications \cite{onnela07}. Figure \ref{fig:example_network} shows a fraction of the social network centered on a single user. The resulting dataset contains the mobility trajectories of 6 million users and a call graph of 33 million edges.

\subsection{Measure Definition}
\label{sec:measures}
We introduce two measures of individual mobility behavior and two measures of individual social behavior, dividing them into two categories: measures of volume, and measures of diversity (see Table \ref{tab:measures}).
\begin{table}[htb]\centering
	\def\arraystretch{1.2}
	\begin{tabular}{ l c c }
		\multicolumn{3}{ c }{ \textbf{individual measures} } \\
		\hline
		\multicolumn{1}{ | l  }{\multirow{2}{*}{\textbf{sociality}}} & \multicolumn{1}{ | c }{social volume} & \multicolumn{1}{ | c |}{$SV$} \\
		\multicolumn{1}{ | l  }{} & \multicolumn{1}{ | c }{social diversity} & \multicolumn{1}{ | c  |}{$SD$} \\
		\hline
	        \multicolumn{1}{ | l  }{\multirow{2}{*}{\textbf{mobility}}} & \multicolumn{1}{ | c }{mobility volume} & \multicolumn{1}{ | c |}{$MV$} \\
		\multicolumn{1}{ | l  }{} & \multicolumn{1}{ | c  }{mobility diversity} & \multicolumn{1}{ | c | }{$MD$} \\
		\hline
	         & & \\
	         \vspace{-7mm}\\
	         \multicolumn{3}{c}{\textbf{socio-economic indicators}}\\	
		\hline
		\multicolumn{1}{ | c  }{\textbf{demographic}} & \multicolumn{1}{ | c  }{population density} & \multicolumn{1}{ | c |  }{$PD$}\\
		\hline
		\multicolumn{1}{ | l  }{\multirow{2}{*}{\textbf{development}}} & \multicolumn{1}{ | c  }{deprivation index} & \multicolumn{1}{ | c | }{$DI$} \\
		\multicolumn{1}{ | l  }{} & \multicolumn{1}{ | c  }{per capita income} & \multicolumn{1}{ | c|  }{$PCI$}\\
		\hline
	\end{tabular}
	\label{tab:measures}
	\caption{Measures and indicators used in our study. Social volume, social diversity, mobility volume and mobility diversity are individual measures computed on mobile phone data. Population density, deprivation index, and per capita income are external socio-economic indicators provided by INSEE.}
\end{table}

We define two measures that capture aspects of individual social interactions: \emph{social volume} ($SV$), the number of social contacts of an individual; and \emph{social diversity} ($SD$), the diversification of an individual's calls over the social contacts. 
Within a social network, we can express the volume of social interactions by counting the amount of links an individual possesses with others. This simple measure of connectivity is widely used in network science and is called the \emph{degree} of an individual \cite{newman03}. In a call graph the degree of an individual is the number of different individuals who are in contact by mobile phone calls with her. We can therefore see the degree as a proxy for the volume of sociality for each individual:
\begin{equation}
SV(u)= degree(u)
\end{equation}
The degree distribution is well approximated by a power law function denoting a high heterogeneity in social networks with respect to the number of friendships \cite{Leskovec2008,onnela07}.

The social diversity of an individual $u$ quantifies the topological diversity in a social network as the Shannon entropy associated with her communication behavior \cite{eagle}:
\begin{equation}
SD(u)= - { \sum_{v=1}^k p_{uv} \log(p_{uv}) \over \log(k)}
\end{equation}
where $k$ is the degree of individual $u$, $p_{uv}= {V_{uv}\over \sum_{v=1}^k V_{uv}}$ and $V_{uv}$ is the number of calls between individual $u$ and individual $v$ during the period of observation. $SD$ is a measure for the social diversification of each individual according to its own interaction pattern. In a more general way, individuals who always call the same few contacts reveal a low social diversification resulting in lower values for $SD$, whereas individuals who distribute their call among many different contacts show high social diversification, i.e.\ higher $SD$. The distribution of $SD$ across the population is peaked, as measured in GSM and landlines data \cite{eagle}.

\begin{figure}[htb]\centering
\subfigure[]{\resizebox*{4.1cm}{!}{\includegraphics{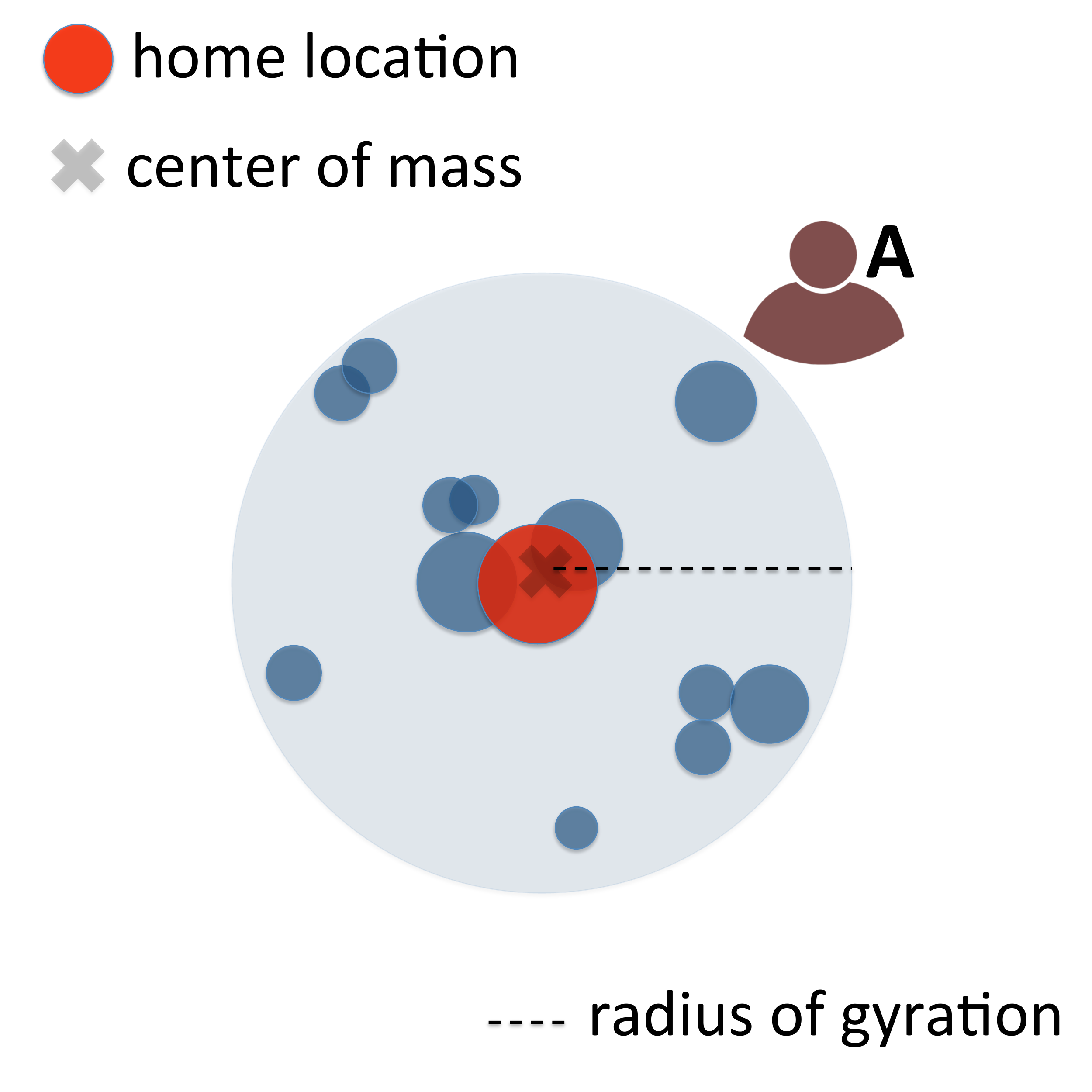}}}
\subfigure[]{\resizebox*{4.1cm}{!}{\includegraphics{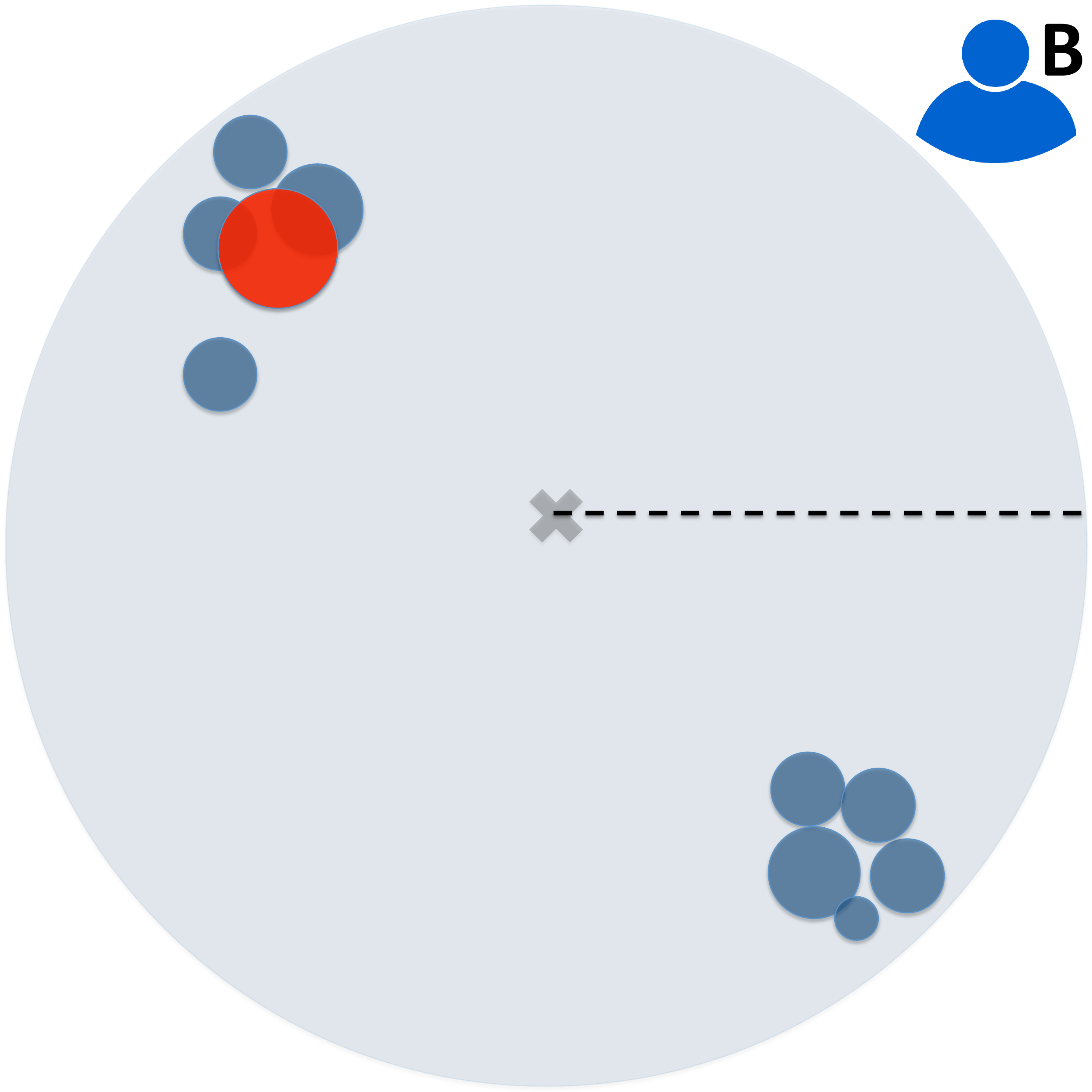}}}
\caption{\textbf{The radius of gyration of two users in our dataset}. The figure shows the spatial distribution of phone towers (circles). The size of circles is proportional to their visitation frequency, the red location indicates the most frequent location $L_1$ (the location where the user makes the highest number of calls during nighttime). The cross indicates the position of the center of mass, the black dashed line indicates the radius of gyration. User $A$ has a small radius of gyration because she travels between locations that are close to each other. User $B$ has high radius of gyration because the locations she visits are far apart from each other.}
\label{fig:examples_volume}
\end{figure}

Starting from the mobility trajectories of an individual, we define two measures to describe individual mobility: \emph{mobility volume} ($MV$), the typical travel distance of an individual, and \emph{mobility diversity} ($MD$), the diversification of an individual's movements over her locations. The radius of gyration \cite{gonzalez08} provides with a measure of mobility volume, indicating the characteristic distance traveled by an individual (see Figure \ref{fig:examples_volume}). In detail, it characterizes the spatial spread of the phone towers visited by an individual $u$ from the trajectories' center of mass (i.e.\ the weighted mean point of the phone towers visited by an individual), defined as:
\begin{equation}
MV(u)= \sqrt{{1\over N} \sum_{i\in L} n_i (r_i - r_{cm})^2}
\end{equation}
where $L$ is the set of phone towers visited by the individual $u$, $n_i$ is the individual's visitation frequency of phone tower $i$, $N= \sum_{i\in L} n_i$ is the sum of all the single frequencies, $r_i$ and $r_{cm}$ are the vectors of coordinates of phone tower $i$ and center of mass respectively. It is known that the distribution of the radius of gyration reveals heterogeneity across the population: most individuals travel within a short radius of gyration but others cover long distances on a regular basis, as measured on GSM and GPS data \cite{gonzalez08,pappalardo2013}.

\begin{figure}[htb]\centering
\subfigure[]{\resizebox*{4.1cm}{!}{\includegraphics{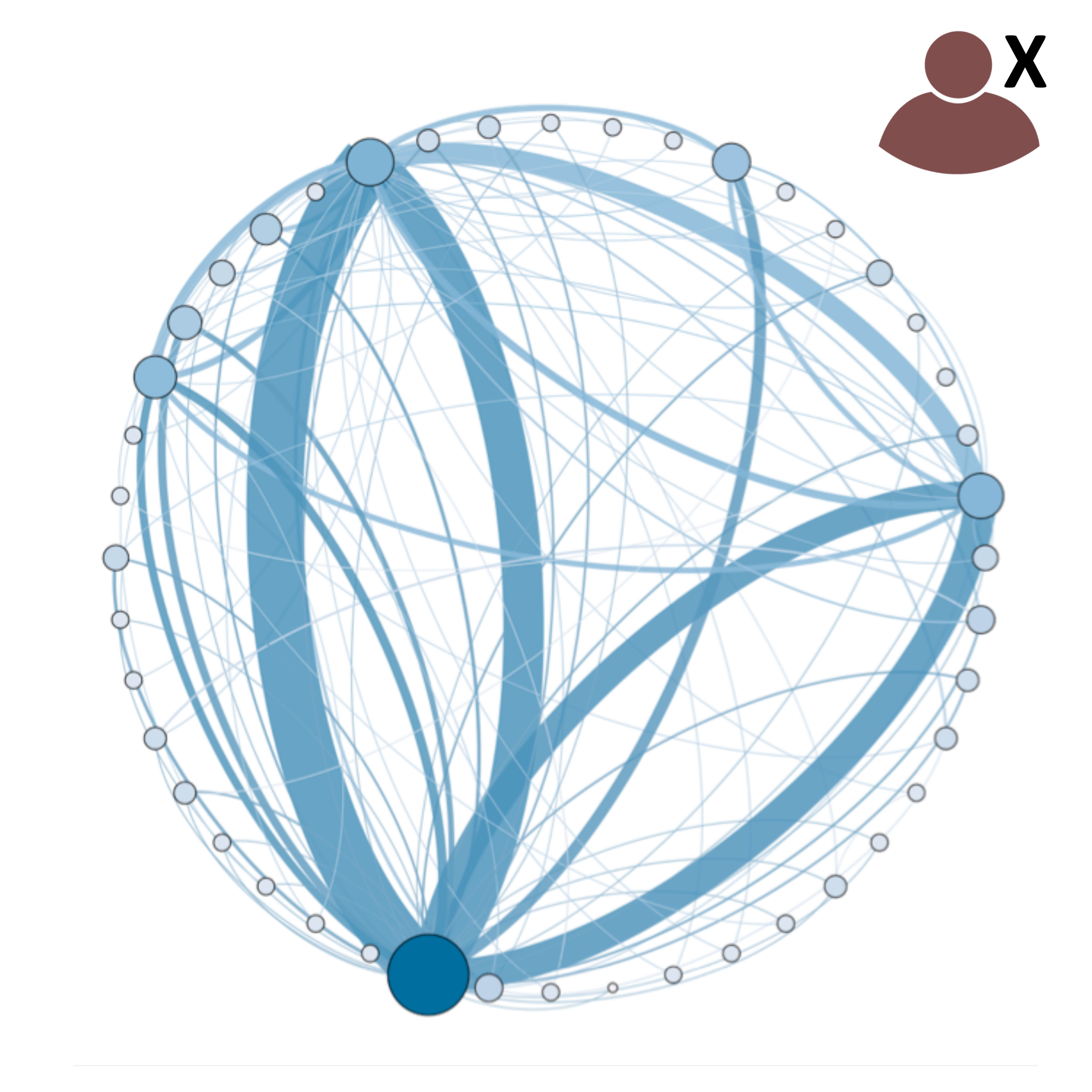}}}
\subfigure[]{\resizebox*{4.1cm}{!}{\includegraphics{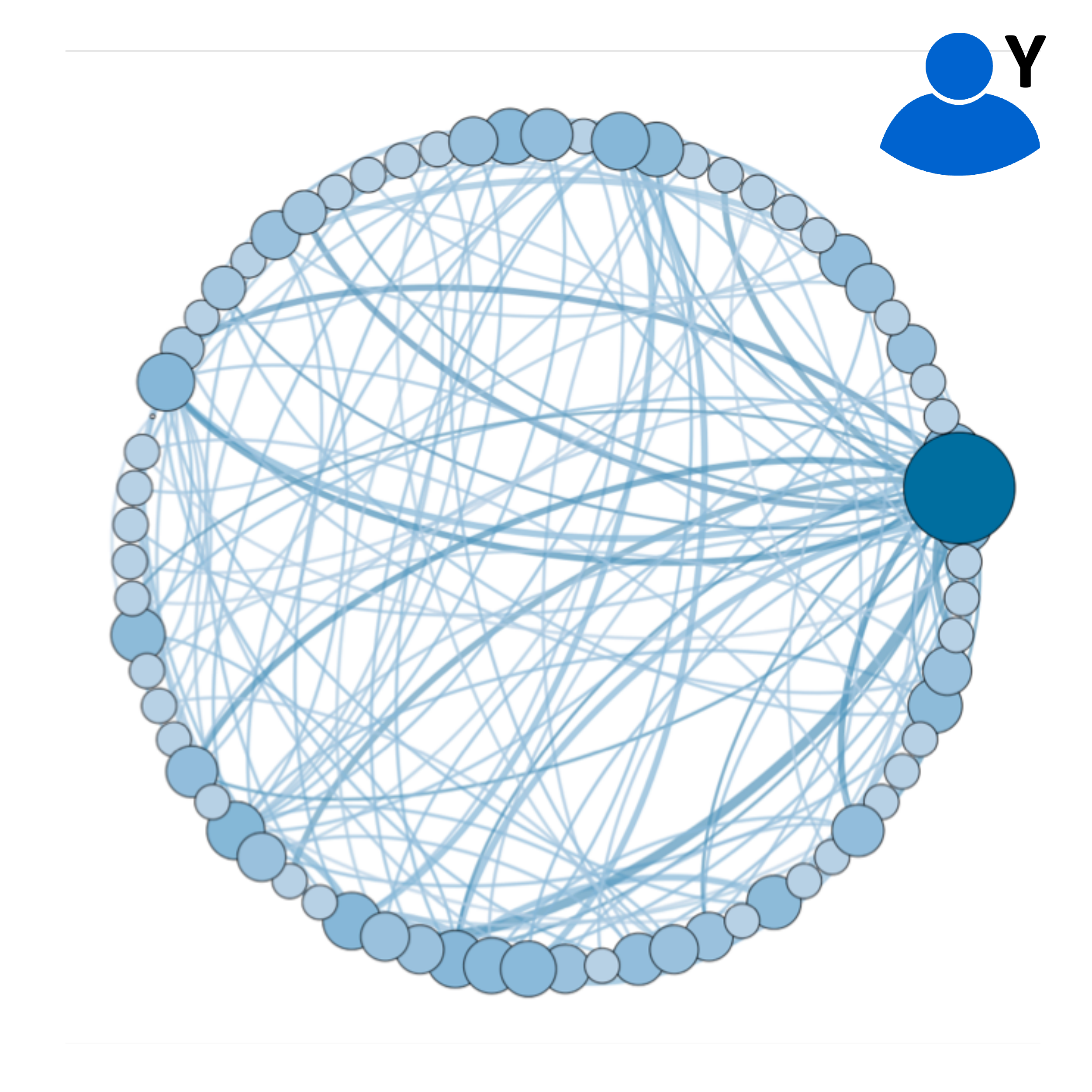}}}
\caption{\textbf{The mobility entropy of two users in our dataset.} Nodes represent phone towers, edges represent trips between two phone towers, the size of nodes indicates the number of calls of the user managed by the phone tower, the size of edges indicates the number of trips performed by the user on the edge. User $X$ has low mobility entropy because she distributes the trips on a few large preferred edges. User $Y$ has high mobility entropy because she distributes the trips across many equal-sized edges.}
\label{fig:examples_diversity}
\end{figure}

Besides the volume of individual mobility, we define the diversity of individual mobility by using Shannon entropy of individual's trips:
\begin{equation}
MD(u)= - {\sum_{e \in E} p(e) \log p(e) \over \log N}
\end{equation}
where $e=(a,b)$ represents a trip between an origin phone tower and a destination phone tower, $E$ is the set of all the possible origin-destination pairs, $p(e)$ is the probability of observing a movement between phone towers $a$ and $b$, and $N$ is the total number of trajectories of individual $u$ (Figure \ref{fig:examples_diversity}). Analogously to $SD$, $MD$ is high when a user performs many different trips from a variety of origins and destinations; $MD$ is low when a user performs a small number of recurring trips. Seen from another perspective, the mobility diversity of an individual also quantifies the possibility to predict individual's future whereabouts. Individuals having a very regular movement pattern possess a mobility diversity close to zero and their whereabouts are rather predictable. Conversely, individuals with a high mobility diversity are less predictable. It is known that the distribution of the mobility diversity is peaked across the population and very stable across different social groups (e.g.\ age and gender) \cite{song2010limits}. 

\subsection{Measure computation}
\label{sec:computation}
We implement step (b) in Figure \ref{fig:process} by computing the four behavioral measures for each individual on the filtered CDR data. Due the size of the dataset, we use the MapReduce paradigm implemented by Hadoop to distribute the computation across a cluster of coordinated nodes and reduce the time of computation.
We find no relationship between the mobility and the social measures at individual level: the correlation between $SV$ and $MV$, as well as the correlation between the $SD$ and the $MD$, are close to zero. This suggests that the mobility measures and the sociality measures capture different aspects of individual behavior.
%\begin{figure}[htb!]
%\begin{center}
%	\resizebox*{4.15cm}{!}{\includegraphics{img/mobility_volume_vs_social_volume.png}}
%	\resizebox*{4.15cm}{!}{\includegraphics{img/mobility_diversity_vs_social_diversity.png}}
%\caption{The relation between (a) mobility volume and social volume; (b) mobility diversity and social diversity. We find no significant correlation between the measures, suggesting that individual mobility and individual sociality capture different aspects of individual behavior.}
%\label{fig:scatterplot_municipalities}
%\end{center}
%\end{figure}

\begin{figure}[htb!]
\begin{center}
	%\resizebox*{8.8cm}{!}{\includegraphics{img/france_orange_population_log10_min_max.png}}
	\resizebox*{8.8cm}{!}{\includegraphics{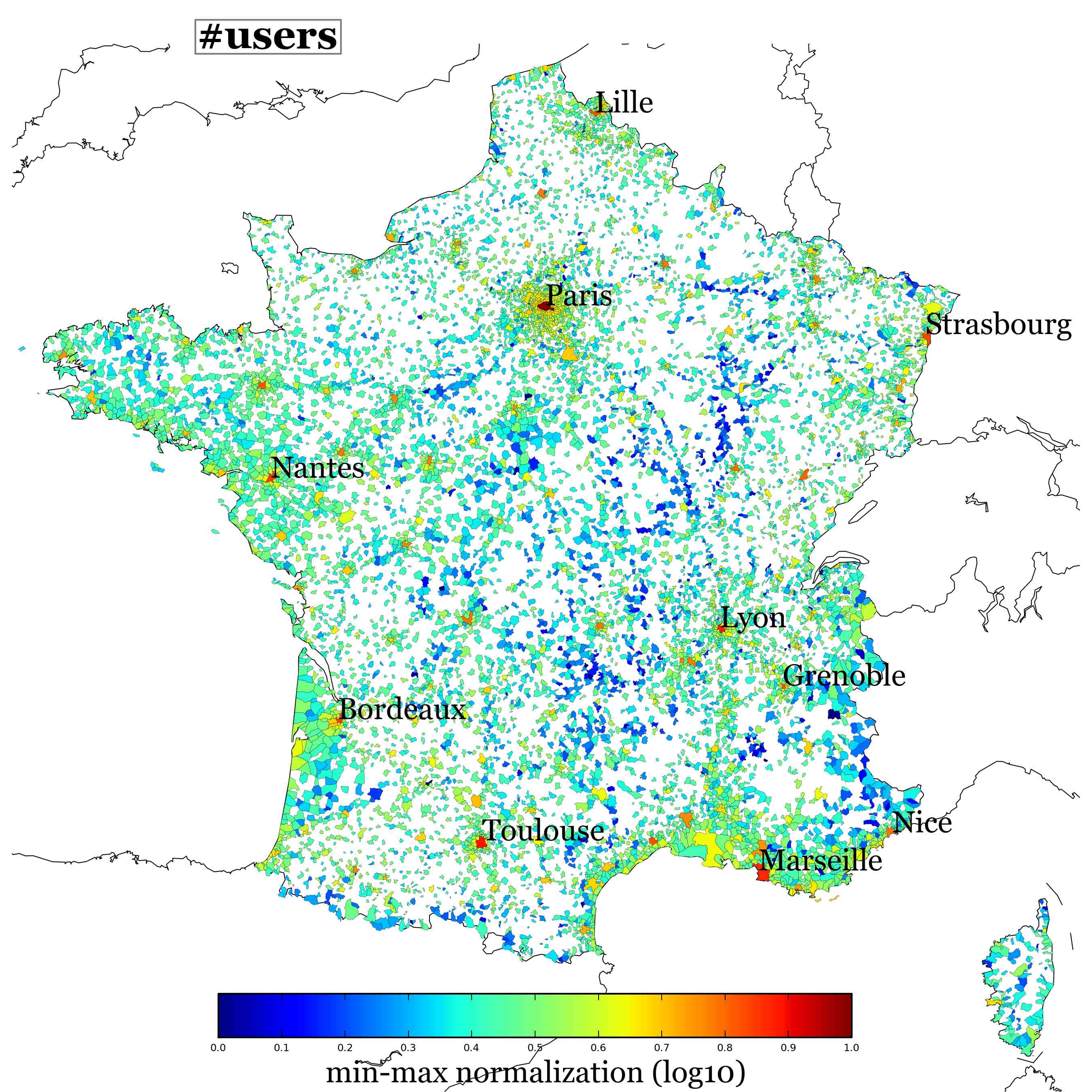}}
\caption{The spatial distribution of users over French municipalities with more than 1,000 official residents. Each user is assigned to a municipality according to the geographic position of her home location. The color of municipalities, in a gradient  from blue to red, indicates the number of Orange users assigned to that municipality. We observe that the number of users in the municipalities varies according to the density of the municipality. }
\label{fig:france}
\end{center}
\end{figure}

We apply step (c) in Figure \ref{fig:process} by aggregating the individual measures at the municipality level through a two-step process: (i) we assign to each user a home location, i.e.\ the phone tower where the user performs the highest number of calls during nighttime (from 10 pm to 7 am) \cite{smoreda2012}; (ii) based on these home locations, we assign each user to the corresponding municipality with standard Geographic Information Systems techniques. Figure \ref{fig:france} shows the spatial distribution of Orange users in French municipalities.
We aggregate the $SV$, $SD$, $MV$ and $MD$ at municipality level by taking the mean values across the population of users assigned to that municipality. We obtain $5,100$ municipalities each one with the associated four aggregated measures.

\section{Correlation Analysis}
\label{sec:analysis}
Here we realize step (d) in Figure \ref{fig:process} and study the interplay between human mobility, social interactions and socio-economic development at municipality level. First, in Section \ref{sec:correlations} we introduce the external socio-economic indicators and investigate their correlation between the behavioral measures aggregated at municipality level. Then in Section \ref{sec:null_models} we compare the results with two null models to reject the hypothesis that the correlations appear by chance. 

\subsection{Human Behavior versus Socio-Economic Development}
\label{sec:correlations}
As external socio-economic indicators, we use a dataset provided by the French National Institute of Statistics and Economic Studies (INSEE) about socio-economic indicators for all the French municipalities with more than 1,000 official residents. 
We collect data on population density ($PD$), per capita income ($PCI$), and a deprivation index ($DI$) constructed by selecting fundamental needs associated both with objective and subjective poverty \cite{pornet2012}. The deprivation index is constructed by selecting among variables reflecting individual experience of deprivation: the different variables are combined into a single score by a linear combination with specific choices for coefficients (see Appendix). Therefore deprivation index is a composite index: the higher its value, the lower is the well-being of the municipality.
Preliminary validation showed a high association between the French deprivation index and both income values and education level in French municipalities, partly supporting its ability to measure socio-economic development \cite{pornet2012}.

\begin{figure*}[htb!]
\begin{center}
\subfigure[]{\resizebox*{4.25cm}{!}{\includegraphics{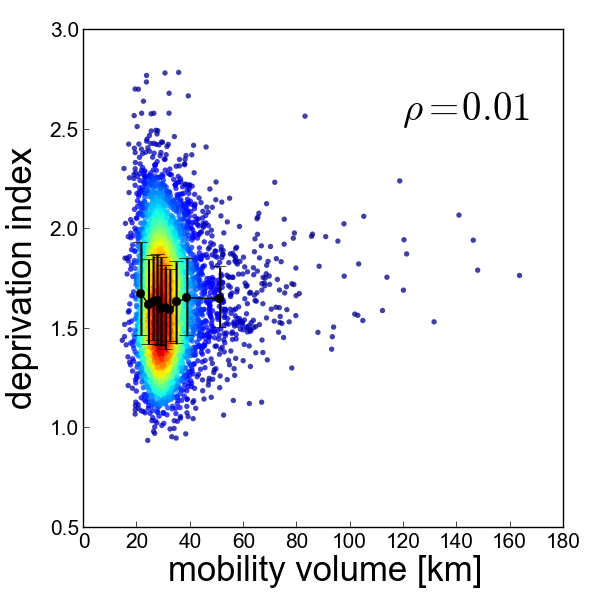}}}
	\subfigure[]{\resizebox*{4.25cm}{!}{\includegraphics{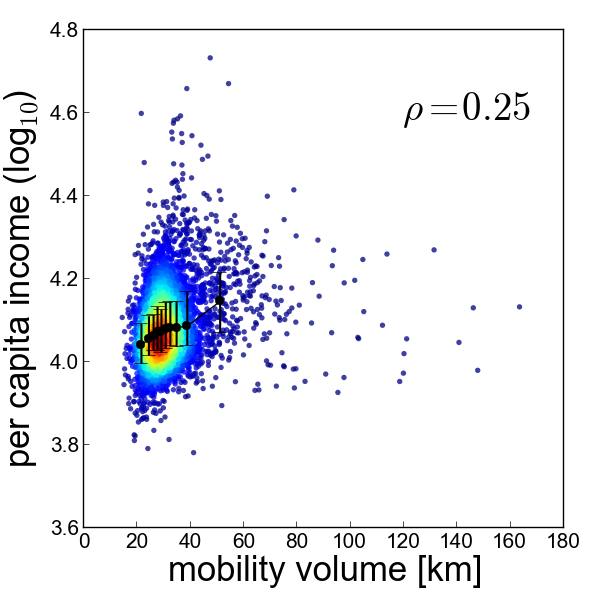}}}
	\subfigure[]{\resizebox*{4.25cm}{!}{\includegraphics{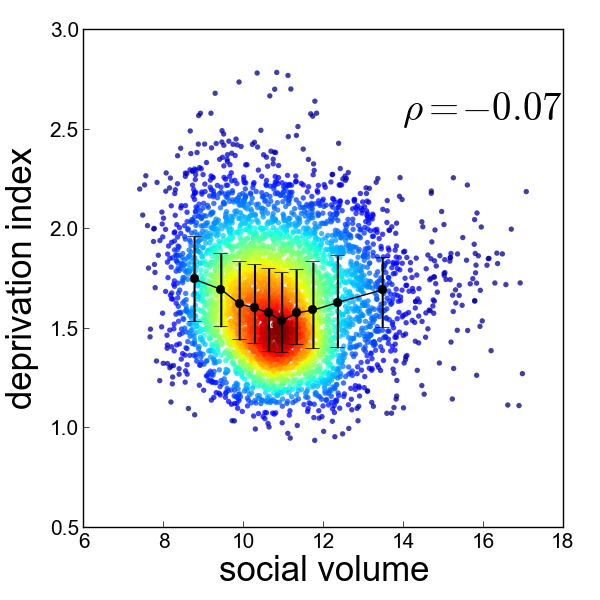}}}
	\subfigure[]{\resizebox*{4.25cm}{!}{\includegraphics{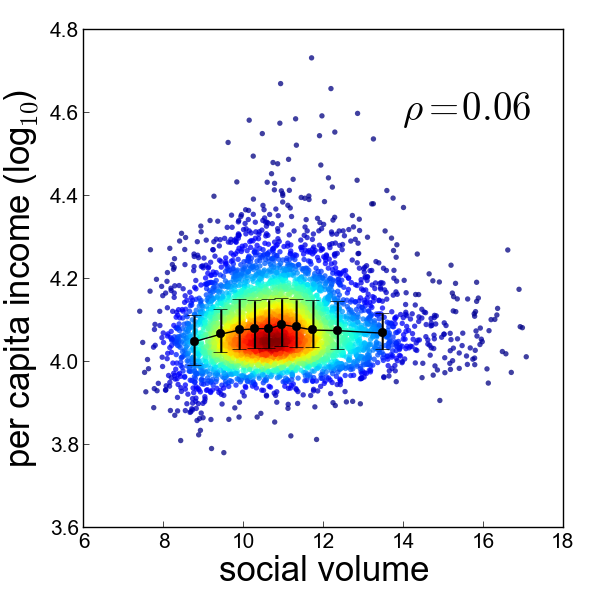}}}
	\subfigure[]{\resizebox*{4.25cm}{!}{\includegraphics{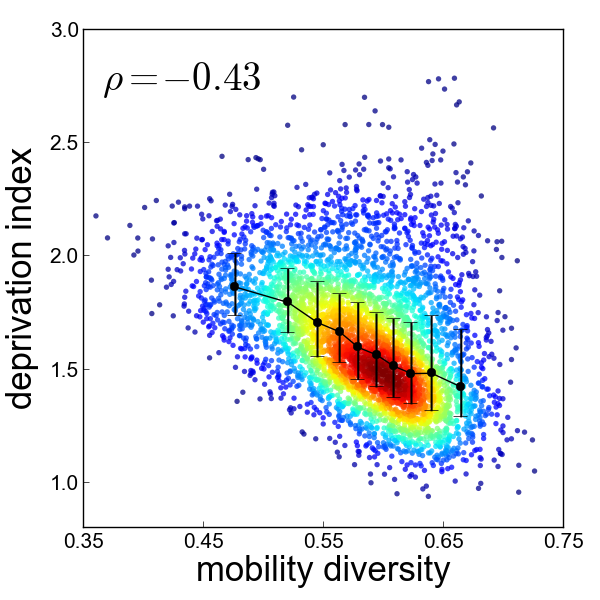}}}
	\subfigure[]{\resizebox*{4.25cm}{!}{\includegraphics{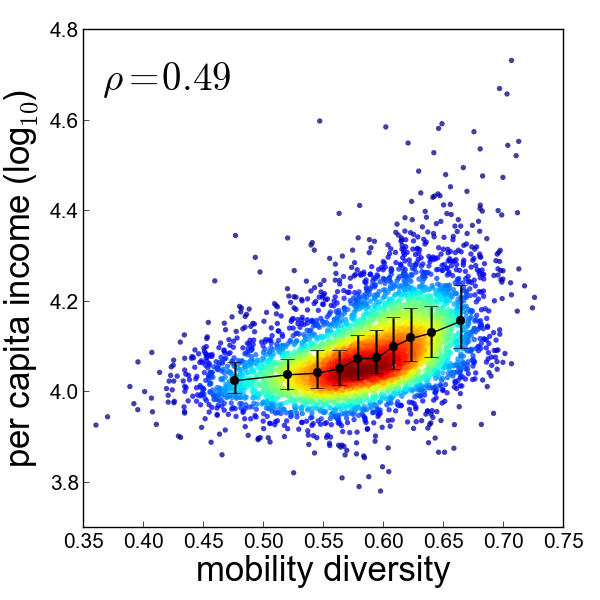}}}
	\subfigure[]{\resizebox*{4.25cm}{!}{\includegraphics{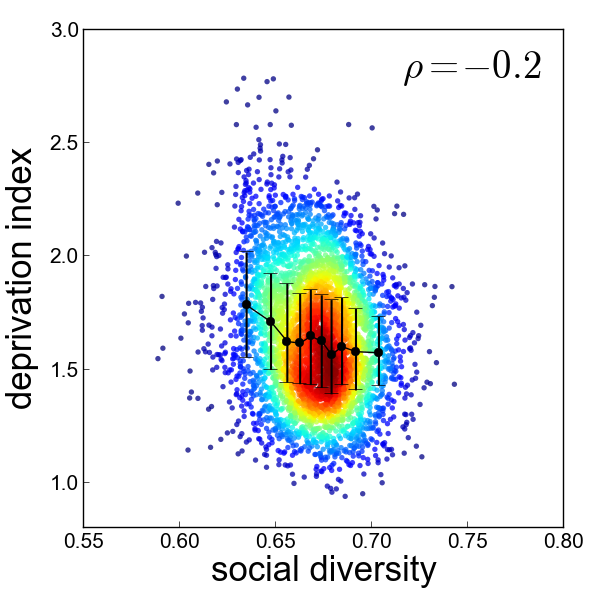}}}
	\subfigure[]{\resizebox*{4.25cm}{!}{\includegraphics{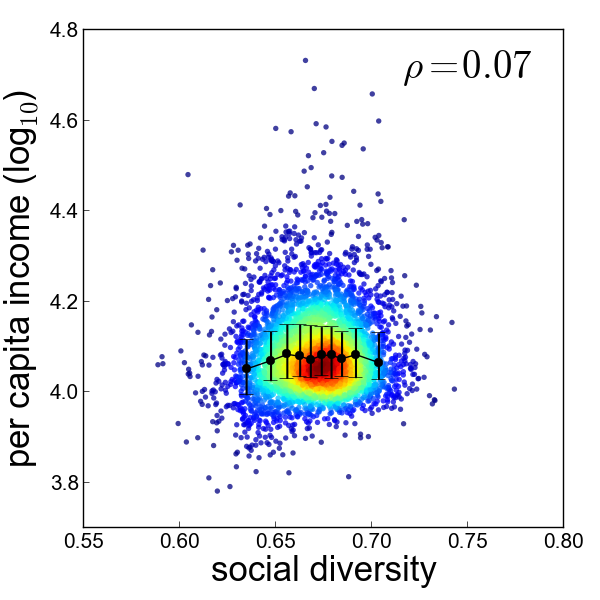}}}
\caption{The relation between the aggregated diversity measures and the socio-economic indicators: (a) mobility volume vs deprivation index; (b) mobility volume vs per capita income; (b) social volume vs deprivation index; (d) social volume vs per capita income; (e) mobility diversity vs deprivation index; (f) mobility diversity vs per capita income; (g) social diversity vs deprivation index; (h) social diversity vs per capita income. The color of a point indicates, in a gradient from blue to red, the density of points around it. We split the municipalities into ten equal-sized groups according to the deciles of the measures on the x axis. For each group, we compute the mean and the standard deviation of the measures on the y axis and plot them through the black error bars. $\rho$ indicates the Pearson correlation coefficient between the two measures. In all the cases the p-value of the correlations is $< 0.001$. }
\label{fig:correlations}
\end{center}
\end{figure*}

We investigate the correlations between the aggregated measures and the external socio-economic indicators finding two main results. First, the social volume is not correlated with the two socio-economic indicators (Figure \ref{fig:correlations}(c) and (d)), while mobility volume is correlated with per capita income (Figure \ref{fig:correlations}(b)). Second, we find that mobility diversity is a better predictor for socio-economic development than social diversity. Figure \ref{fig:correlations}(e)-(h) shows the relations between diversity measures and socio-economic indicators. For mobility diversity clear tendencies appear: as the mean mobility diversity of municipalities increases, deprivation index decreases, while per capita income increases (Figure \ref{fig:correlations}(e) and (f)). Social diversity, in contrast, exhibits a weaker correlation with the deprivation index than mobility diversity and no correlation with per capita income (Figure \ref{fig:correlations}(g) and (h)).

\begin{figure}[htb]
\begin{center}
	\subfigure[]{\resizebox*{4cm}{!}{\includegraphics{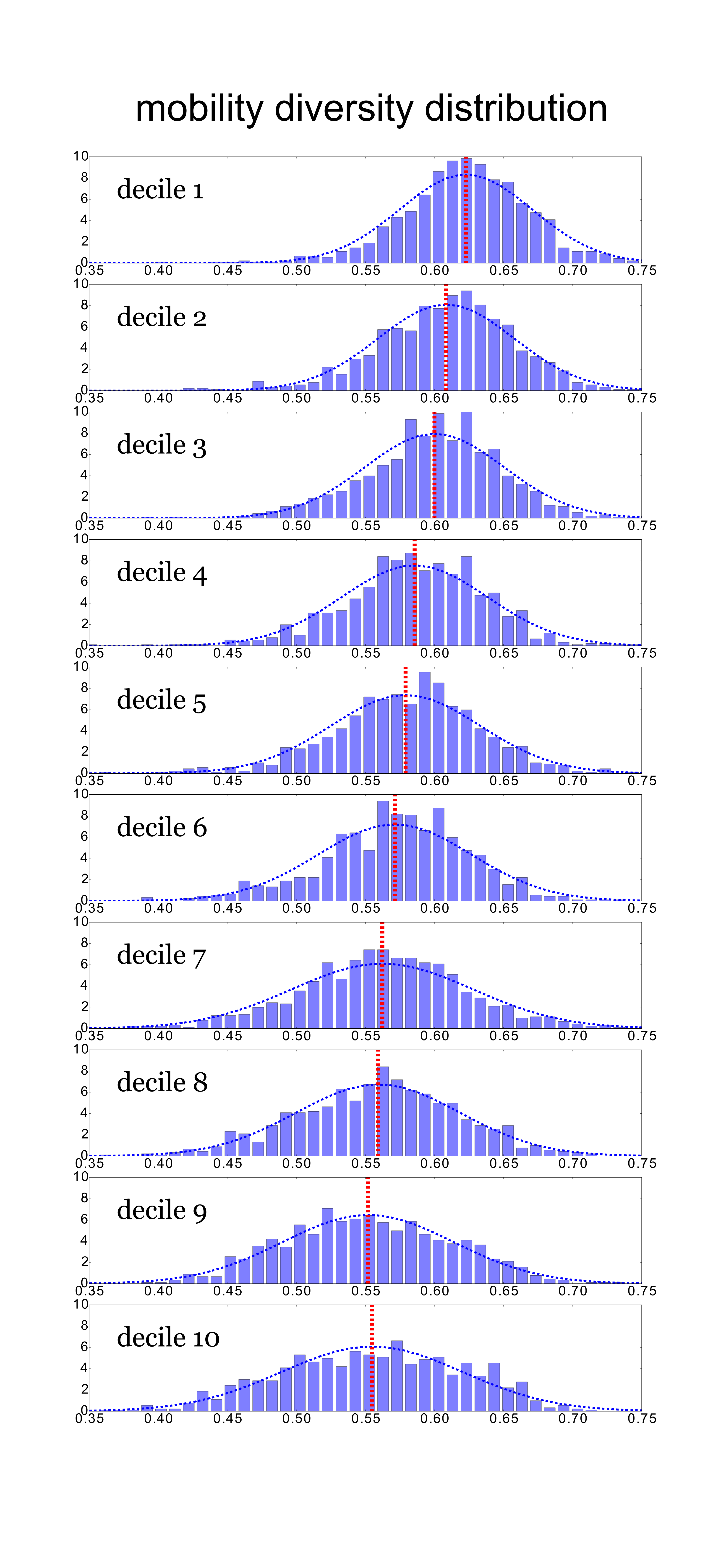}}}
	\subfigure[]{\resizebox*{4cm}{!}{\includegraphics{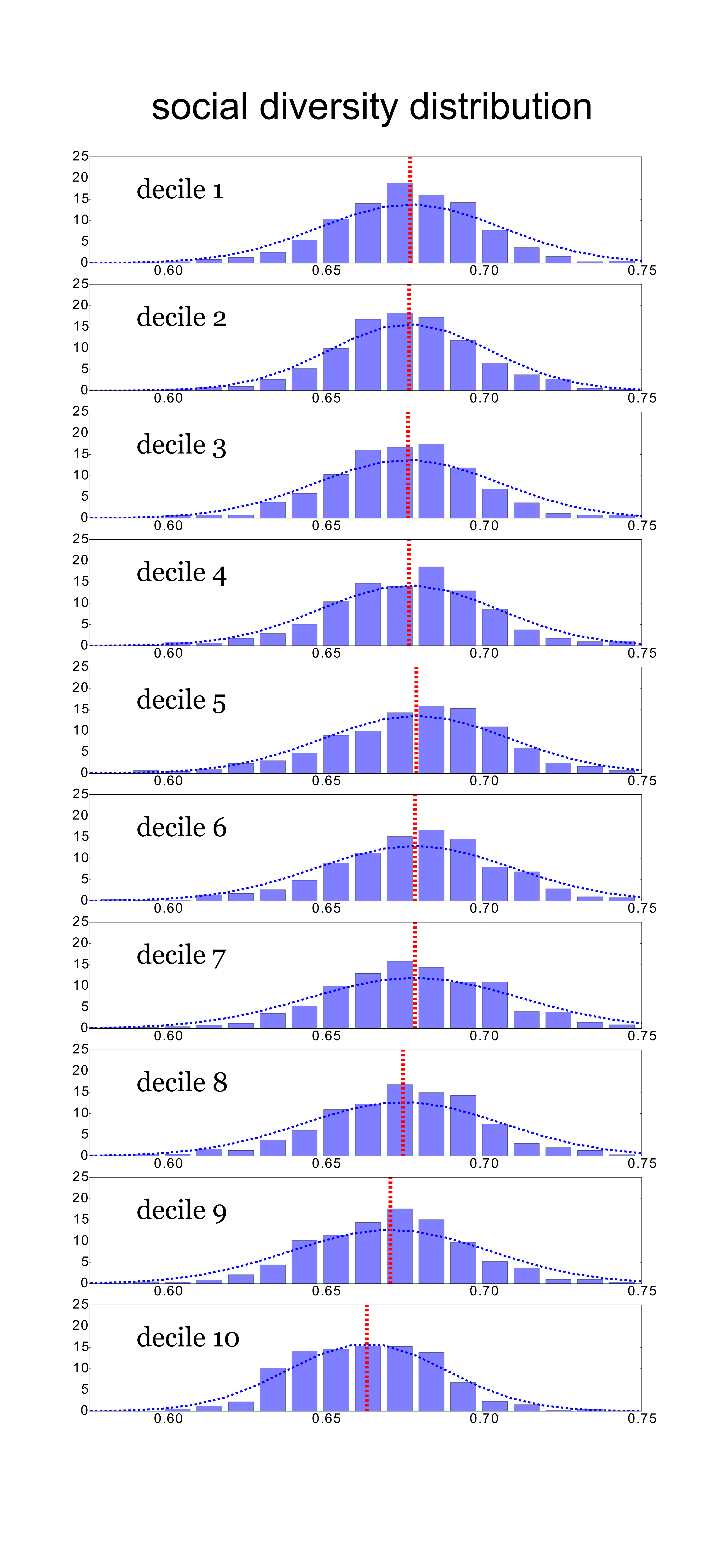}}}
\caption{The distribution of mobility diversity (a) and social diversity (b) in the deciles of deprivation index. We split the municipalities into ten equal-sized groups computed according to the deciles of deprivation index. For each group, we plot the distributions of mean mobility diversity and mean social diversity. The blue dashed curve represent a fit of the distribution, the red dashed line represents the mean of the distribution.}
\label{fig:distributions}
\end{center}
\end{figure}

Figure \ref{fig:distributions} provides another way to observe the relations between the diversity measures and socio-economic development. We split the municipalities in ten deciles according to the values of deprivation index. For each decile we compute the distributions of mean mobility diversity and mean social diversity across the municipalities in that decile. For mobility diversity, the deciles of the economic values increase while the mean decreases and the variance increases, highlighting a change of the distribution in the different groups. This is consistent with the observation made in the plots of Figure \ref{fig:correlations}(e). Conversely, for social diversity distribution we do not observe a significant change in the mean and the variance. The observed variation of the mobility diversity distribution in the different deciles is an interesting finding when compared to previous works such as Song et al.\  \cite{SongNaturePhysics2010} which states that mobile predictability is very stable across different subpopulations delineated by personal characteristics like gender or age group.
Figure \ref{fig:correlations} and \ref{fig:distributions} suggest us that the diversity of human mobility aggregated at municipality level is better associated with the socio-economic indicators than socio-demographic characteristics.
The relation between mobility diversity and deprivation index is stronger and more evenly distributed over the different levels of deprivation index for municipalities.

%\vspace{-1mm}
%\begin{mdframed}[linewidth=1pt]
%\setlength{\parindent}{0pt}
%Mobility diversity shows the strongest correlations\\ with socio-economic indicators.
%\end{mdframed}

\subsection{Validation against Null Models}
\label{sec:null_models}
In order to test the significance of the correlations observed on the empirical data, we compare our findings with the results produced by two null models. 

In null model NM1, we randomly distribute the users over the French municipalities. We first extract uniformly $N$ users from the dataset and assign them to a random municipality with a population of $N$ users. We then aggregate the individual diversity measures of the users assigned to the same municipality. We repeat the process 100 times and take the mean of the aggregated values of each municipality produced in the 100 experiments. 

\begin{figure*}[htb!]
\begin{center}
	\subfigure[]{\resizebox*{4.25cm}{!}{\includegraphics{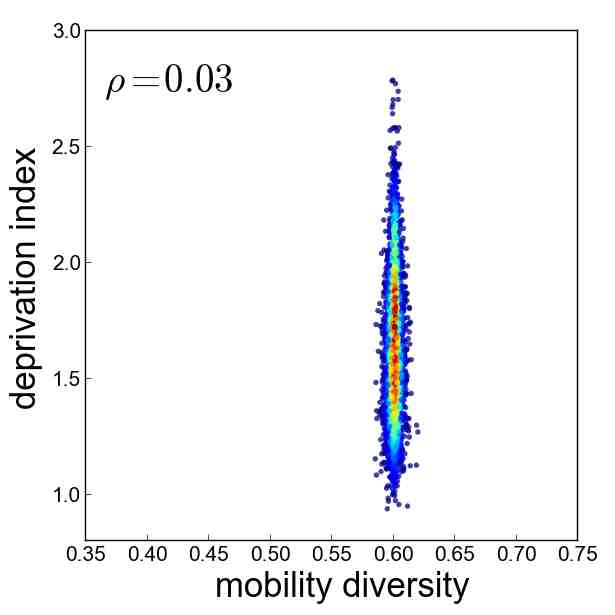}}}
	\subfigure[]{\resizebox*{4.25cm}{!}{\includegraphics{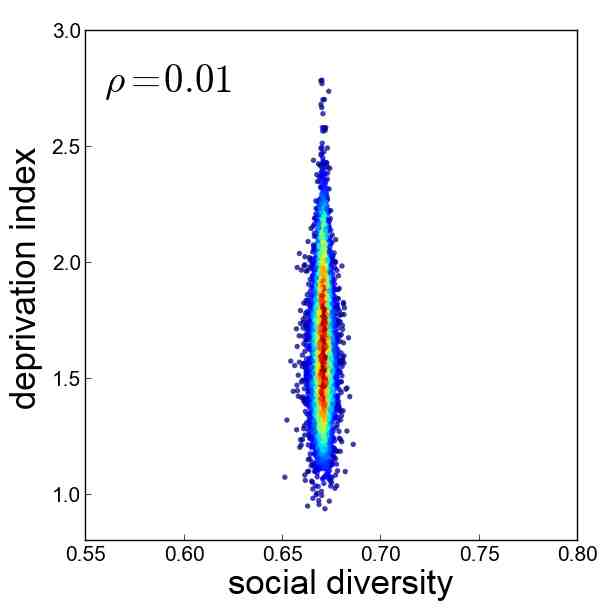}}}
	\subfigure[]{\resizebox*{4.25cm}{!}{\includegraphics{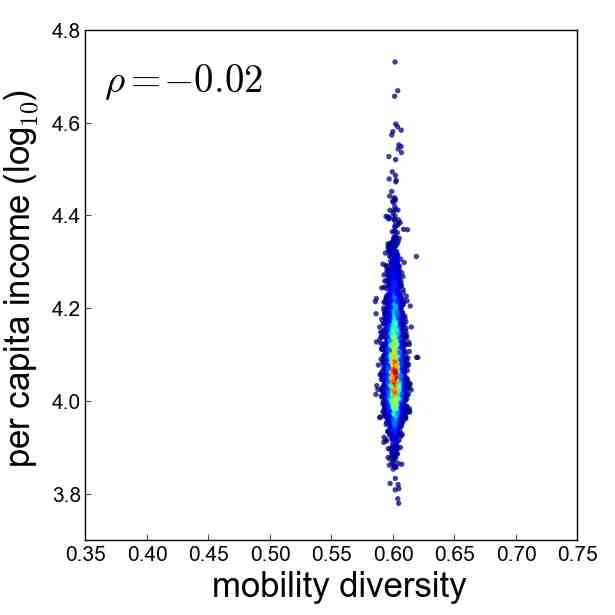}}}
	\subfigure[]{\resizebox*{4.25cm}{!}{\includegraphics{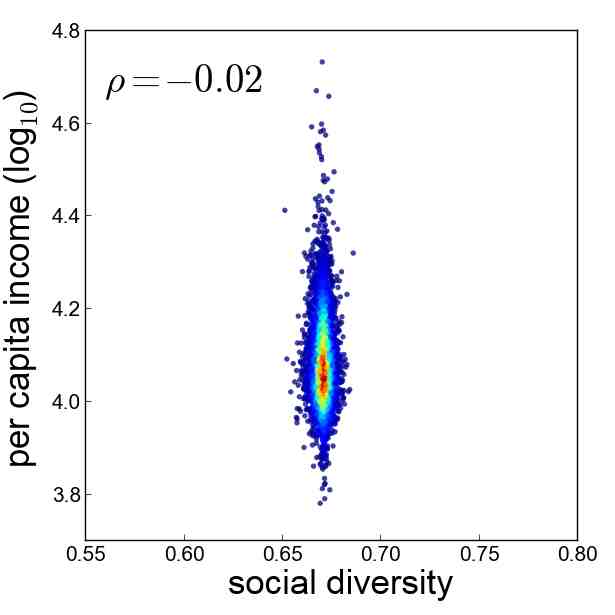}}}
		\subfigure[]{\resizebox*{4.25cm}{!}{\includegraphics{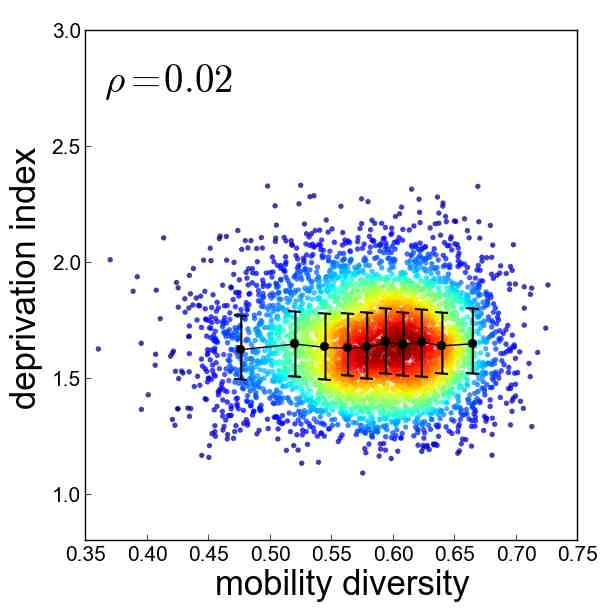}}}
	\subfigure[]{\resizebox*{4.25cm}{!}{\includegraphics{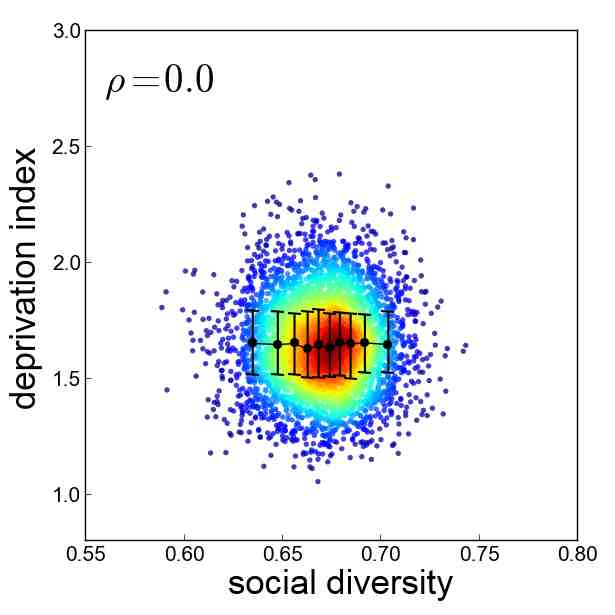}}}
	\subfigure[]{\resizebox*{4.25cm}{!}{\includegraphics{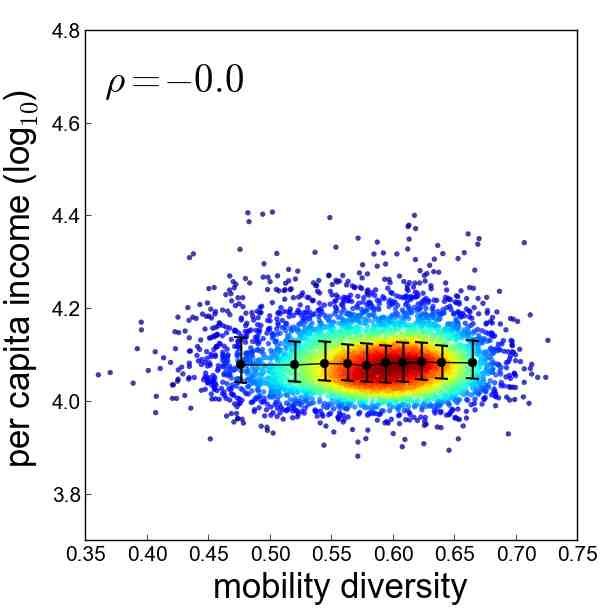}}}
	\subfigure[]{\resizebox*{4.25cm}{!}{\includegraphics{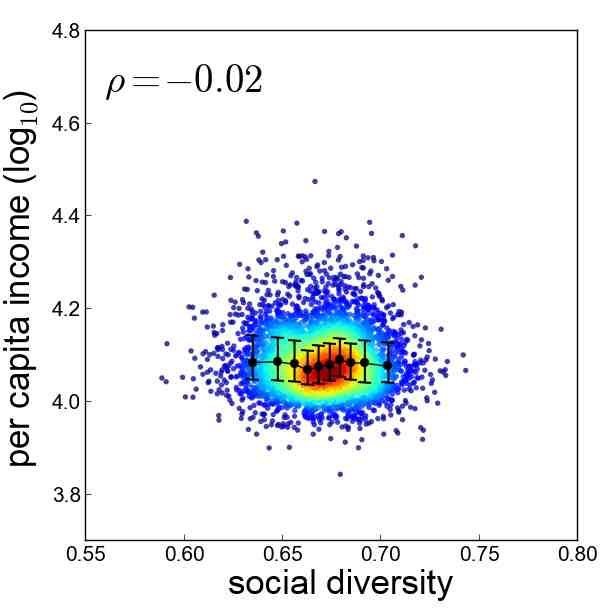}}}
\caption{The relation between the socio-economic indicators and the diversity measures computed on null model NM1 (a-d) and null model NM2 (e-h). The color of a point indicates, in a gradient from blue to red, the density of points around it. We split the municipalities into ten equal-sized groups according to the deciles of the measures on the x axis. For each group, we compute the mean and the standard deviation of the measure on the y axis (the black error bars).}
\label{fig:correlations_null_models}
\end{center}
\end{figure*}

In null model NM2, we randomly shuffle the values of the socio-economic indicators over the municipalities. We perform this procedure 100 times and take, for each municipality, the mean value of the socio-economic indicators computed over the 100 produced values. In contrast with empirical data, we find no correlation in the null models between the diversity measures and the socio-economic indicators, neither for mobility diversity nor for social diversity (Figure \ref{fig:correlations_null_models}). Such a clear difference between the correlations observed over empirical data and the absence of correlations in observations on randomized data allows us to reject the hypothesis that our findings are obtained by chance.

%\vspace{-4.5mm}
%\begin{mdframed}[linewidth=1pt]
%Null Models produce zero correlations between behavioral measures and socio-economic indicators.
%\end{mdframed}
%Figure \ref{fig:summary} summarizes our findings. Here we organize municipalities by deciles based on the deprivation index. We then compute the mean values for both mobility diversity and social diversity of these decile groups. The results confirm our initial interpretation. The associations between the deprivation index and the mobility diversity are more explicit then the associations between the deprivation index and the social diversity. Generalizing, we conclude that human mobility proves to be a better predictor for the economic development of cities when compared to social behavior.
%\begin{figure}[htb!]
%\begin{center}
%	\subfigure[]{\resizebox*{4.15cm}{!}{\includegraphics{img/mobility_diversity_decile_deprivation_index.png}}}
%	\subfigure[]{\resizebox*{4.15cm}{!}{\includegraphics{img/social_diversity_decile_deprivation_index.png}}}
%\caption{The mean values of mobility diversity (a) and social diversity (b) for groups of municipalities based on the deciles of the deprivation index. We split municipalities into ten equal-sized groups according to the deciles of the deprivation index. For each group, we compute the mean for the mobility diversity and the social diversity.}
%\label{fig:summary}
%\end{center}
%\end{figure}

\section{Predictive Models}
\label{sec:models}
In this section we instantiate step (e) building and validating both regression models (Section \ref{sec:regressions}) and classification models (Section \ref{sec:classifications}) to predict the external socio-economic indicators from the aggregated measures. 

\subsection{Regression Models}
\label{sec:regressions}
To learn more about the relationship between the aggregated measures and the socio-economic indicators we implement two multiple regression models M1 and M2. We use deprivation index as dependent variable in model M1, per capita income as dependent variable for model M2, the four aggregated measures and population density as regressors for both models. We determine the regression line using the least squared method. The model M1 for deprivation index produces a coefficient of determination $R^2 = 0.43$, meaning that the regressors explain the 43\% of the variation in the deprivation index. The model M2 for per capita income explains the 25\% of the variation in the per capita income producing a a coefficient of determination $R^2=0.25$. Table \ref{tab:models} and Table \ref{tab:model_M2} show the coefficients of the regression equations, the standard error of the coefficients and the p-values of the regressors for model M1 and model M2 respectively. For both model M1 and M2 we have verified the absence of multicollinearity between the regressors, the normality and the homoskedasticity of regression residuals.
\begin{table}[htb]
\begin{tabular}{llllll}
\multicolumn{4}{c}{\qquad \textbf{Model M1} (deprivation index), $R^2 = 0.4267$} \\
\hline
 & \textbf{coefficients}   & \textbf{std. error} & \textbf{p-value} &  \\
\textbf{PD}  & 0.247 & 0.005 & $< 2\times10^{-16}$ \\
\textbf{MD} &    -2.980 & 0.0575 & $< 2\times10^{-16}$ \\
\textbf{SD} &  -2.153 & 0.2027 & $< 2\times10^{-16}$\\
\textbf{MV} &    0.002 & 0.0002  & $5.35\times10^{-16}$ \\
\textbf{SV} &     0.006 & 0.0027 &   $ 0.013$  \\
\textbf{intercept}  & 4.078 & 0.1281 & $< 2\times10^{-16}$  \\
\hline
\end{tabular}
\caption{The linear regression model M1 for deprivation index. The \emph{coefficients} column specifies the value of slope calculated by the regression. The \emph{std. error} column measures the variability in the estimate for the coefficients. The \emph{p-value} column shows the probability the variable is not relevant.}
\label{tab:models}
\end{table}
\begin{table}[htb]\centering
\begin{tabular}{llllll}
\multicolumn{4}{c}{\qquad \textbf{Model M2} (per capita income), $R^2=0.25$} \\
\hline
 & \textbf{coefficients}   & \textbf{std. error} & \textbf{p-value} &  \\
\textbf{PD}  & 781.94 & 74.84 & $< 2\times10^{-16}$ \\
\textbf{MD} & 22,773.47 & 729.05 & $< 2\times10^{-16}$ \\
\textbf{SD} &  18,451.79 & 2,569.05 & $7.82\times10^{-13}$\\
\textbf{MV} &    63.116 & 3.64  & $<2\times10^{-16}$ \\
\textbf{SV} &     191.16 & 34.62 &   $ 3.56\times10^{-8}$  \\
\textbf{intercept}  & -18,933.66 & 1,624.36 & $< 2\times10^{-16}$  \\
\hline
\end{tabular}
\caption{The linear regression model M2 for per capita income. The \emph{coefficients} column specifies the value of slope calculated by the regression. The \emph{std. error} column measures the variability in the estimate for the coefficient. The \emph{p-value} column shows the probability the variable is not relevant. }
\label{tab:model_M2}
\end{table}

\begin{figure}[htb]
\begin{center}
	\subfigure[]{\resizebox*{4.15cm}{!}{\includegraphics{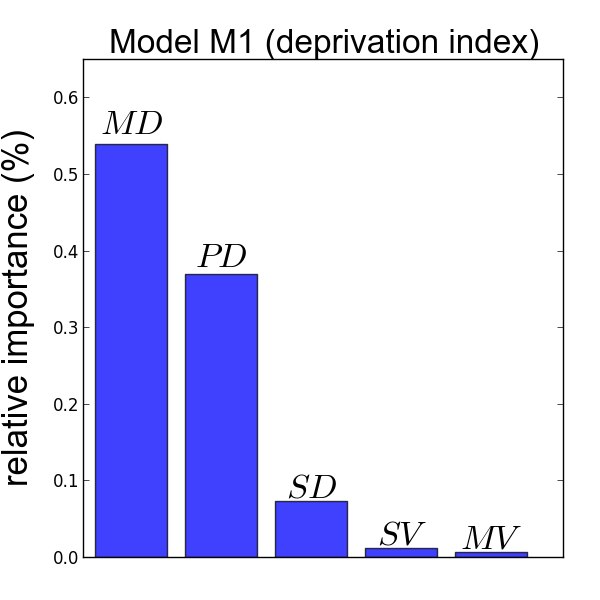}}}
	\subfigure[]{\resizebox*{4.15cm}{!}{\includegraphics{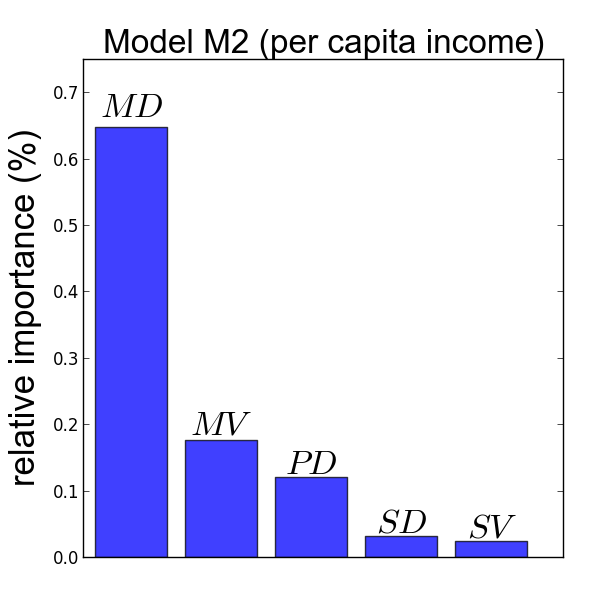}}}
\caption{The relative importance of the aggregated measures in the multiple regression models M1 (a) and M2 (b). We use the Lindeman, Merenda and Gold (LMG) method to quantify an individual regressor's contribution to the model. We observe that mobility diversity is the most important variable in the model with a contribution of about 54\% and 65\% for model M1 and M2 respectively.}
\label{fig:relative_importance}
\end{center}
\end{figure}

We quantify the contribution of each regressor to the multiple regression model by computing a relative importance metric \cite{groemping}. Figure \ref{fig:relative_importance} shows the relative importance of regressors produced by the LMG method \cite{lindeman1980} for both model M1 and model M2. We observe that mobility diversity gives the highest contribution to the regression, accounting for the 54\% and 65\% of the importance for M1 and M2 respectively, while social diversity provides a little contribution (0.7\% for M1 and 0.3\% for M2). Population density provides an important contribution in both models, mobility volume is an important variable to model M2 only (20\% of the variance).

\begin{figure}[htb]
\begin{center}
	\subfigure[]{\resizebox*{4.15cm}{!}{\includegraphics{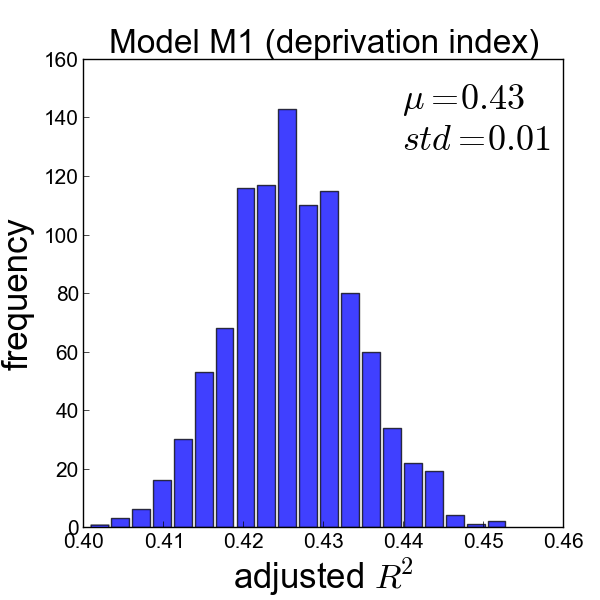}}}
	\subfigure[]{\resizebox*{4.15cm}{!}{\includegraphics{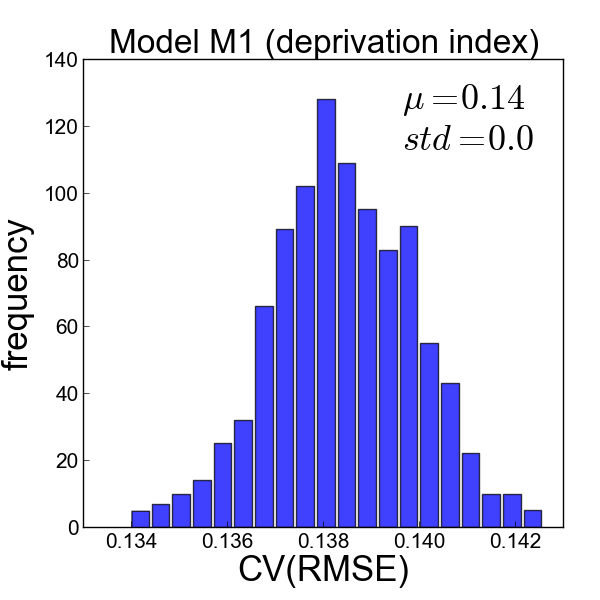}}}
	\subfigure[]{\resizebox*{4.15cm}{!}{\includegraphics{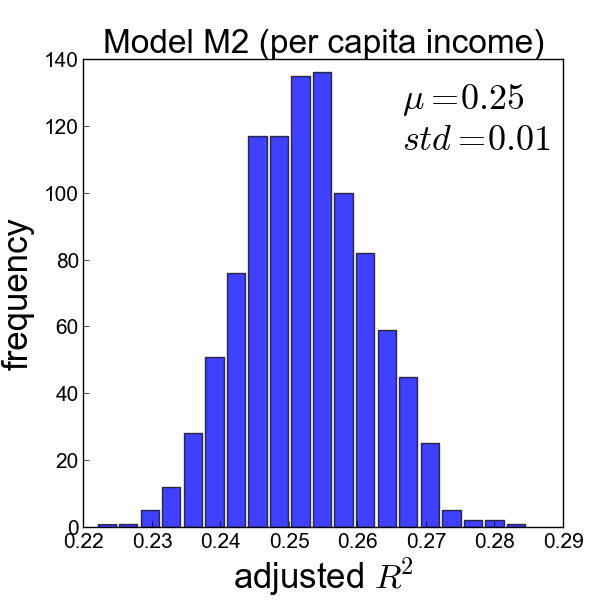}}}
	\subfigure[]{\resizebox*{4.15cm}{!}{\includegraphics{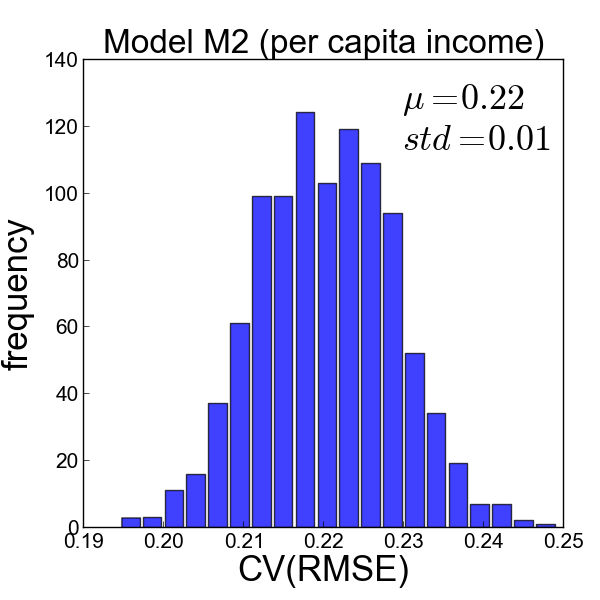}}}
\caption{Validation of regression models. We perform 1,000 experiments learning the model on a training set (60\%) and evaluating it on a a test set (40\%). (a) The distribution of the adjusted coefficient of determination $R^2$ across the experiments for model M1. (b) The distribution of the root mean square error (RMSE) across the experiments for model M1. (c) The distribution of the adjusted $R^2$ across the experiments for model M2. (d) The distribution of RMSE for model M2. }
\label{fig:errors}
\end{center}
\end{figure}

\begin{figure}[htb]
\begin{center}
	\subfigure[]{\resizebox*{4.15cm}{!}{\includegraphics{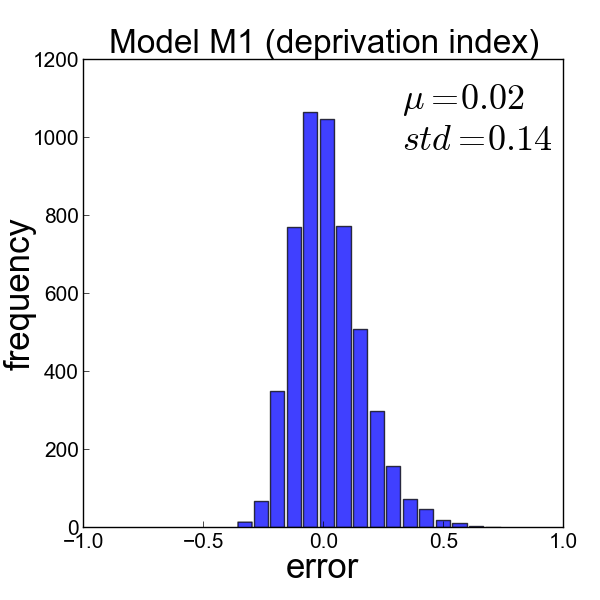}}}
	\subfigure[]{\resizebox*{4.15cm}{!}{\includegraphics{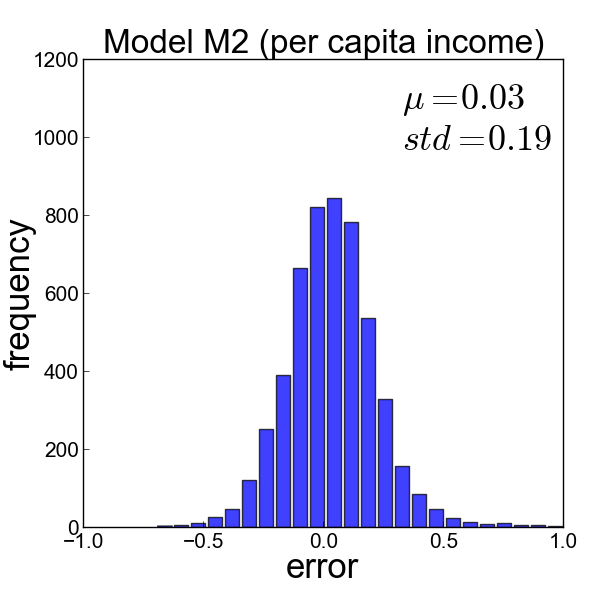}}}
\caption{The distribution of the relative error $(\hat{y}_i - y)/y$ across the French municipalities for regression models M1 (a) and M2 (b).}
\label{fig:errors_cities}
\end{center}
\end{figure}

To validate the models we implement a cross validation procedure by performing 1,000 experiments. In each experiment we divide the dataset of municipalities into a training set (60\%) and a test set (40\%), compute model M1 and model M2 on the training set, and apply the obtained models on the test set. We evaluate the performance of the models on the test set using the root mean square error $RMSE=\sqrt{\sum_i^n (\hat{y}_i - y_i)^2) / n}$, where $\hat{y}_i$ is the value predicted by the model and $ y_i$ the actual value in the test set, and computing the CV(RMSE), i.e.\ the RMSE normalized to the mean of the observed values. Figure \ref{fig:errors} shows the variation of $R^2$ and $CV(RMSE)$ across the 1,000 experiments. We observe that the prediction error of the models is stable across the experiments (Figure \ref{fig:errors}(a) and (c)), and that the error in the prediction is lower for model M1 (deprivation index).
Finally, we compare the actual values of socio-economic indicators and the values predicted by the models by computing the relative error, i.e.\ for each municipality $i$ we compute $(\hat{y}_i - y_i) / y$. We observe that the mean relative error computed across the municipalities is close to zero for both model M1 and model M2 (Figure \ref{fig:errors_cities}).
%\vspace{-2mm}
%\begin{mdframed}[linewidth=1pt]
%Mobility diversity gives the highest contribution in regression models to predict socio-economic development.
%\end{mdframed}

\subsection{Classification Models}
\label{sec:classifications}
Here, instead of predicting the value of deprivation or per capita income of municipalities we want to \emph{classify} the level of socio-economic development of municipalities.
To this purpose we build two supervised classifiers C1 and C2 that assign each municipality to one of three possible categories: low level, medium level or high level of deprivation index (classifier C1) or per capita income (classifier C2). To transform the two continuous measures deprivation index and per capita income into discrete variables we partition the range of values using the 33th percentile of the distribution. This produced, for each variable to predict, three equal-populated classes. We perform the classification using Random Forest classifiers on a training set (60\% of the dataset) and validate the results on a test set (40\% of the dataset). Classifier C1 for deprivation index reaches an overall accuracy of 0.61, while the overall accuracy of classifier C2 for per capita income is 0.54, against a random case accuracy of 0.33. Table \ref{tab:classifiers} shows precision, recall and overall accuracy reached by classifier $C_1$ and classifier $C_2$ on the three classes of socio-economic development.

\begin{figure}[htb]
\begin{center}
	\subfigure[]{\resizebox*{4.15cm}{!}{\includegraphics{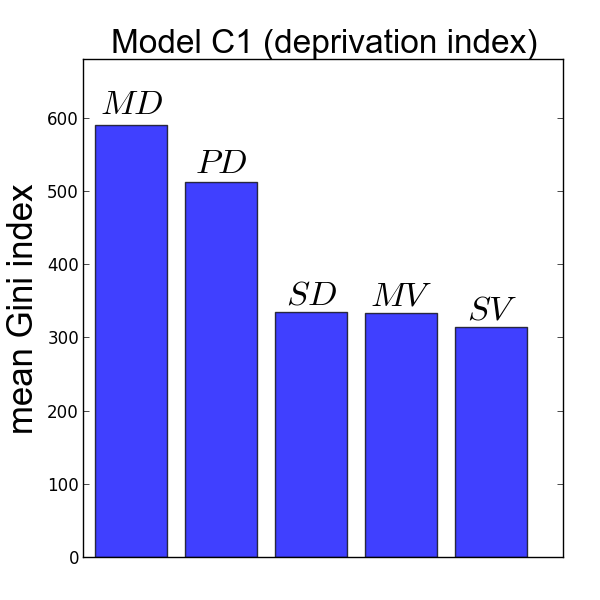}}}
	\subfigure[]{\resizebox*{4.15cm}{!}{\includegraphics{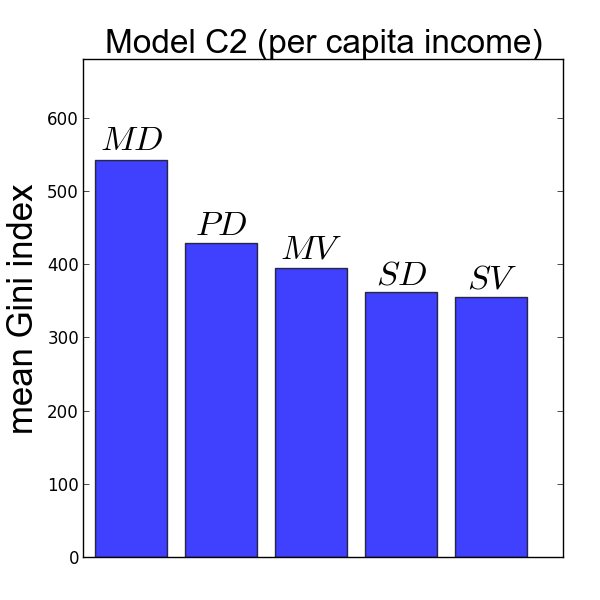}}}
\caption{The mean decrease in Gini coefficient of the variables used to learn the classifiers, for deprivation index (a) and per capita income (b). The mean decrease in Gini coefficient is a measure of how each variable contributes to the homogeneity of nodes and leaves in the resulting random forest.}
\label{fig:gini_index}
\end{center}
\end{figure}

We also evaluate the importance of every aggregated measure in classifying the level of socio-economic development of municipalities, using the Mean Decrease Gini measure. Similarly to the Relative Importance metrics for the regression models, in both classifier $C_1$ and classifier $C_2$ the mobility diversity has the highest importance, followed by population density (Figure \ref{fig:gini_index}).

\begin{table}[htb]\centering
\def\arraystretch{1.2}
\begin{tabular}{ l  c  c  }
\multicolumn{3}{c}{\textbf{Model C1: $accuracy = 0.61$}} \\
\hline
& \textbf{recall} & \textbf{precision}\\
\textbf{low deprivation} & 0.6230 & 0.6657 \\
\textbf{medium deprivation} & 0.4970 & 0.4918 \\
\textbf{high deprivation} & 0.7089 & 0.6721 \\
\hline
\end{tabular}

\bigskip
\smallskip

\begin{tabular}{ l  c  c  }
\multicolumn{3}{c}{\textbf{Model C2}, \emph{accuracy} = 0.54} \\
\hline
& \textbf{recall} & \textbf{precision}\\
\textbf{low income} & 0.6098 & 0.5700 \\
\textbf{medium income} & 0.3590 & 0.3993 \\
\textbf{high income} & 0.6552 & 0.6376 \\
\hline
\end{tabular}

\caption{Statistics by class for classifier C1 (deprivation index) and classifier C2 (per capita income). The recall is the number municipalities for which the classifier predicts the correct class divided by the number of municipalities in that class. The precision is the number of municipalities for which the classifier predicts the correct class divided by the number of municipalities the classifier predicts to be in that class. We observe that the classes `low' and `high' are the best predicted classes.}
\label{tab:classifiers}
\end{table}
%\vspace{-1mm}
%\begin{mdframed}[linewidth=1pt]
%Mobility diversity gives the highest contribution in classification models for socio-economic development.
%\end{mdframed}
%\ \ \\
\section{Discussion of Results}
\label{sec:discussion}
The implementation of the analytical framework on mobile phone data produces three remarkable results. 

First, the usage of the measures of mobility and social behavior together with the standard and commonly available socio-demographic information actually \emph{adds predictive power} with respect to the external socio-economic indicators. Indeed, while a univariate regression that predicts deprivation index from population density is able to explains only the 11\% of the variance, by adding the four behavioral measures extracted from mobile phone data we can explain the 42\% of the variance (see Table \ref{tab:models}). This outcome suggests that mobile phone data are able to provide precise and realistic measurements of the behavior of individuals in their complex social environment, which can be used within a knowledge infrastructure like our analytical framework to monitor socio-economic development.

Second, the \emph{diversification of human movements} is the most important aspect for explaining the socio-economic status of a given territory, far larger than the diversification of social interactions and demographic features like population density. This result, which is evident from both the correlations analysis and the contribution of mobility diversity in the models (Figures \ref{fig:correlations}, \ref{fig:relative_importance} and \ref{fig:gini_index}), is also important for practical reasons. 
%Our models exploit both mobility and social measures extracted from mobile phone data. 
Mobile phone providers do not generally release, for privacy reasons, information about the call interactions between users, i.e.\ the social dimension. Our result shows that this is a marginal problem since the social dimension has a lower impact to the quality of the models than the mobility dimension (Figures \ref{fig:relative_importance} and \ref{fig:gini_index}). Hence, the implementation of our analytical framework guarantees reliable results even when, as often occurs because of privacy and proprietary reasons, the social dimension is not available in the data.

The interpretation of the observed relation between mobility diversity and socio-economic indicators is, without a doubt, two-directed. It might be that a well-developed territory provides for a wide range of activities, an advanced network of public transportation, a higher availability and diversification of jobs, and other elements that foster mobility diversity. As well as it might be that a higher mobility diversification of individuals lead to a higher socio-economic development as it could nourish economy, establish economic opportunities and facilitate flows of people and goods. In any case this information is useful for policy makers, because a change in the diversification of individual movements is linked to a change into the socio-economic status of a territory.

Third remarkable result is that our regression and classification models exhibit good performance when used to predict the socio-economic development of other municipalities, whose data where not used in the learning process (Figure \ref{fig:errors} and Table \ref{tab:classifiers}). This result is evident from the cross validation procedure: the accuracy and the prediction errors of the models are not dependent on the training and test set selected. The models hence give a real possibility to continuously monitor the socio-economic development of territories and provide policy makers with an important tool for decision making.

%From a geographic point of view, our results suggest that scale plays an important role in the relationship between mobility diversity and social diversity. On an individual level, no significant correlations between the two measures exists while at the inter-municipality level a significant relation emerges. The reasons for this scale related emergence of relationships is difficult to assess, and cannot simply be retrieved from our analysis only. Nevertheless, the results suggest that although individual possess their own, unique way of communicating and moving, their environment on a city scale implies some sort of convergence towards the behavior of other individuals in their geographical proximity. The analysis of CDR data form a unique way to empirically describe such scale related processes and might give direction to other future research efforts.

\section{Conclusions and Future works}
\label{sec:conclusions}
In this paper we design an analytical framework that uses mobile phone data to extract meaningful measures of human behavior and estimate indicators for socio-economic development. We apply the analytical framework on a nationwide mobile phone data covering several weeks and find that the diversification of human movements is the best proxy for indicators of socio-economic development. We know that bio-diversity is crucial to the health of natural ecosystems, that the diversity of opinion in a crowd is essential to answer difficult questions \cite{galton} and that the diversity of social contacts is associated to socio-economic
indicators of well-being \cite{eagle}. The story narrated in this paper suggests that diversity is a relevant concept also in mobility ecosystems: the diversity of human mobility may be a reliable indicator of the variety of human activities, and a mirror of some aspects of socio-economic development and well-being. 

We are aware that the computation of individual measures on CDR data (step (a) and (b) in Figure \ref{fig:process}) present privacy issues. An important next step will be to incorporate a privacy-by-design approach. We intend to use a method to assess the privacy risk of users in order to detect risk cases where the privacy of users is violated and apply privacy enhancing techniques for data anonymization \cite{francesca}.

In our experiments we compare the measures of mobility and sociality with two external socio-economic indicators: per capita income and deprivation index. Per capita income is a simple indicator indicating the mean income of individuals resident in a given municipality, without any information about the distribution of the wealth and the inequality. In contrast deprivation index is a composite indicator obtained as linear combination of several different variables regarding economic and ecological aspects (see Appendix). It would be interesting, as future work, to investigate the relation between the behavioral measures and the socio-economic development in a multidimensional perspective, using the single variables composing the deprivation index to understand which are the aspects of socio-economic development that best correlate with the measures of human behavior. This multidimensional approach is fostered by recent academic research and a number of concrete initiatives developed around the world \cite{helbing11,bes14} which state that the measurement of well-being should be based on many different aspects besides the material living standards (income, consumption and wealth): health, education, personal activities, governance, social relationships, environment, and security. All these dimensions shape people's well-being, and yet many of them are missed by conventional income measures. Official statistics institutions are incorporating questions to capture people's life evaluations, hedonic experiences and priorities in their own surveys (see for example the Italian BES project developed by Italian National Statistics Bureau \cite{bes14}). When these measures will become available, they will allow us to refine our study on the relation between measures extracted from Big Data and the socio-economic development of territories.

In the meanwhile, experiences like ours may contribute to shape the discussion on how to measure some of the aspects of socio-economic development with Big Data, such as mobile phone call records, that are massively available everywhere on earth. If we learn how to use such a resource, we have the potential of creating a digital nervous system in support of a generalized and sustainable development of our societies. This is crucial because the decisions policy makers (and we as individual citizens) make depend on what we measure, how good our measurements are and how well our measures are understood.

\section*{Appendix}
\label{ap:deprivation_index}
As described in \cite{pornet2012}, the value of deprivation index for French municipalities is calculated in the following way: 
\begin{equation*}
	\begin{split}
	\mbox{deprivation } = &0.11 \times \mbox{Overcrowding}\\
	 &+ 0.34 \times \mbox{No access to electric heating}\\
	 &+ 0.55 \times \mbox{Non-owner}\\
	 &+ 0.47 \times \mbox{Unemployment}\\
	 &+ 0.23 \times \mbox{Foreign nationality}\\
	 &+ 0.52 \times \mbox{No access to a car}\\
	 &+ 0.37 \times \mbox{Unskilled worker-farm worker}\\
	 &+ 0.45 \times \mbox{Household with 6 + persons}\\
	 &+ 0.19 \times \mbox{Low level of education}\\
	 &+ 0.41 \times \mbox{Single-parent household}.
	\end{split}
\end{equation*}
\begin{figure}[htb]
\begin{center}
	\resizebox*{4cm}{!}{\includegraphics{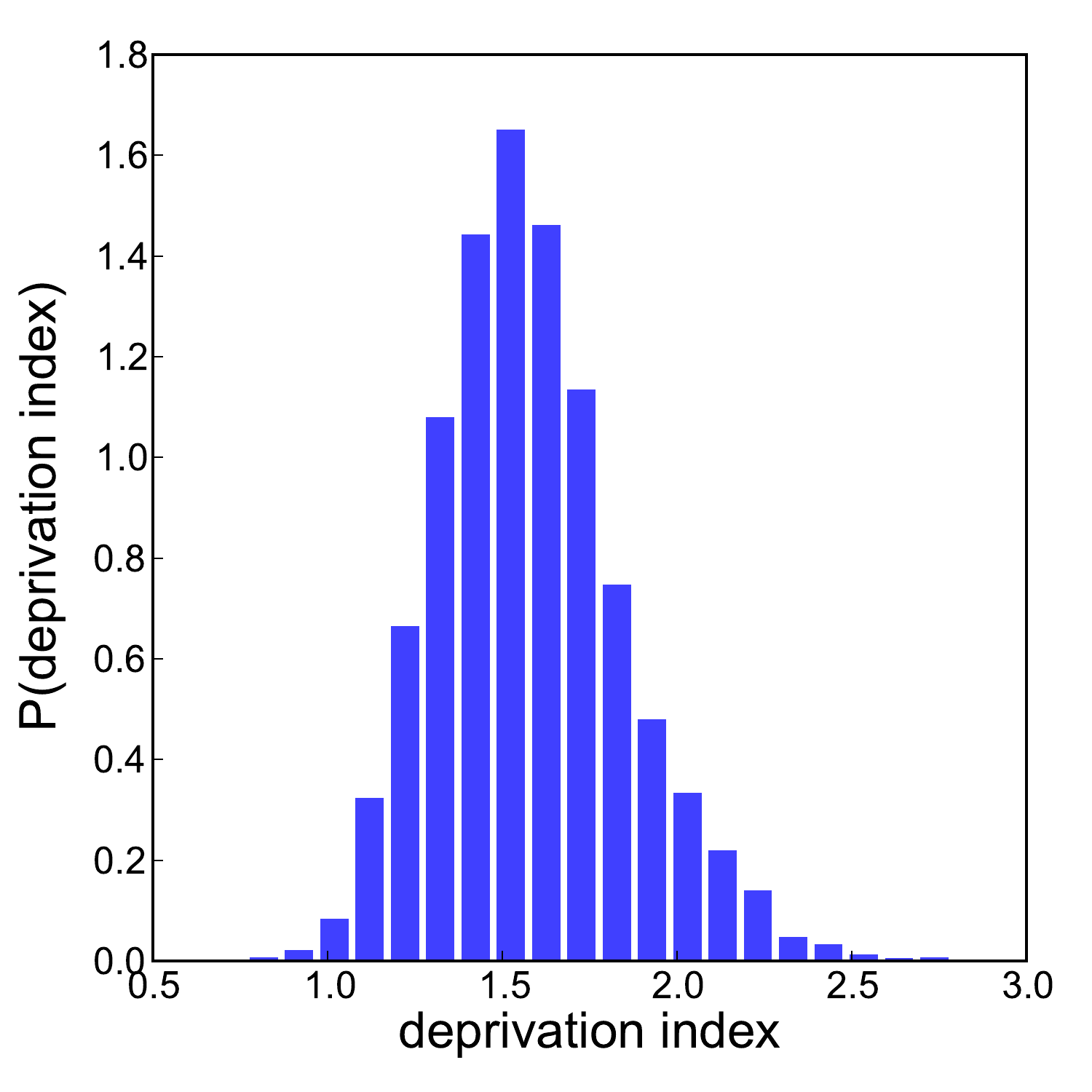}}
	\resizebox*{4cm}{!}{\includegraphics{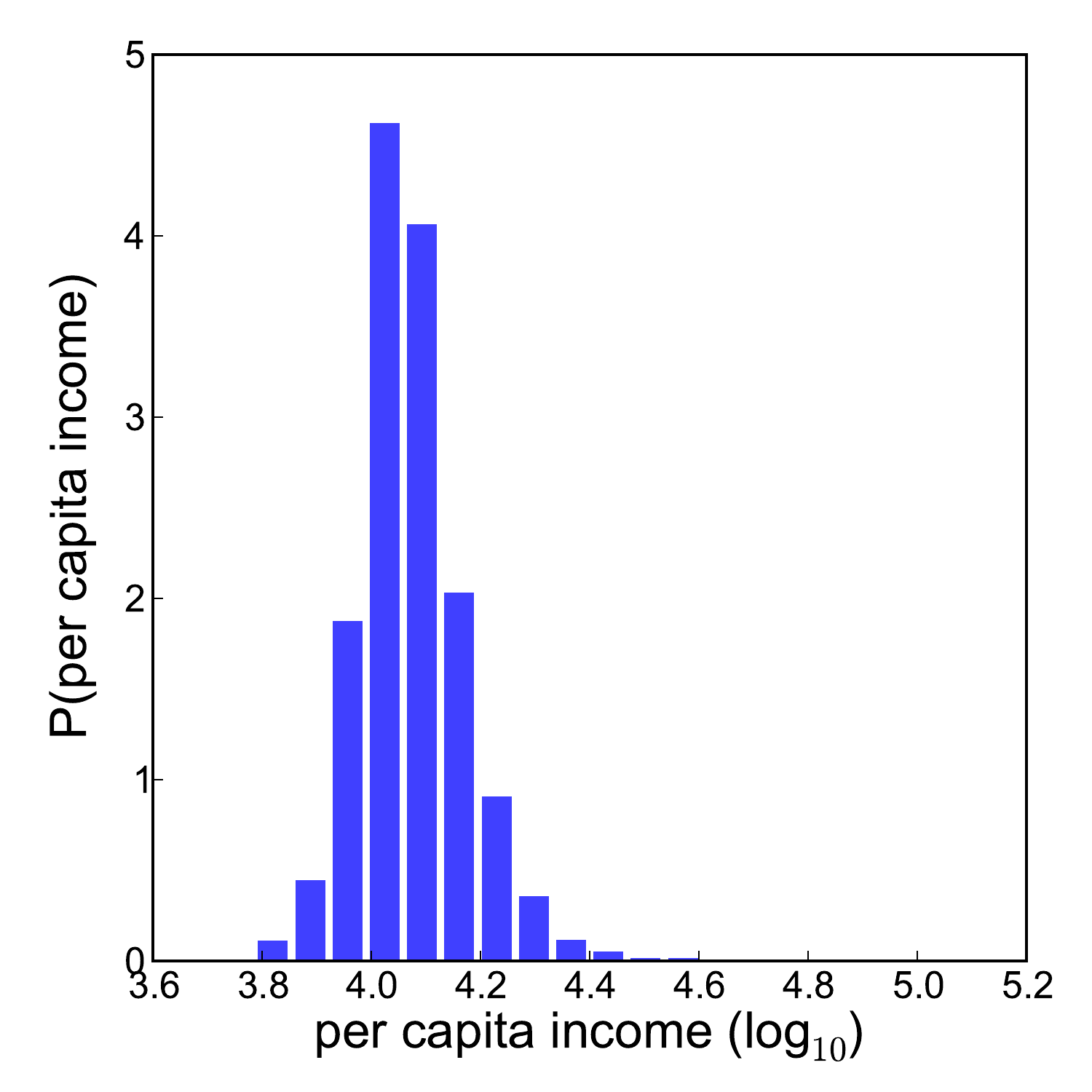}}
\caption{Distribution of deprivation index (a) and per capita income (b) across French municipalities.}
\label{fig:deprivation}
\end{center}
\end{figure}

\acknowledgement{
The authors would like to thank Orange for providing the CDR data, Giovanni Lima and Pierpaolo Paolini for the contribution developed during their master theses. We are grateful to Carole Pornet and colleagues for providing the socio-economic indicators and for computing the deprivation index for the French municipalities. 

This work has been partially funded by the following European projects: Cimplex (grant agreement 641191), PETRA (grant agreement 609042), SoBigData RI (grant agreement 654024). 
}

\bibliographystyle{abbrv}
\bibliography{biblio}

\end{document}